\documentclass[12pt,preprint]{aastex}

\newcommand{\fer}{{\it Fermi}}
\newcommand{\wse}{{\it WISE}}

\newcommand{\gstrp}{WGS}
\usepackage{graphicx}
\usepackage{longtable}
\usepackage[figuresright]{rotating}

\slugcomment{version \today: fm}

\shorttitle{The \wse\ $\gamma$-ray strip parametrization}
\shortauthors{F. Massaro, R. D'Abrusco, G. Tosti, M. Ajello, D. Gasparrini, J. E. Grindlay \& H. A. Smith 2011}

\begin{document}
\title{The \wse\ gamma-ray strip parametrization: \\ the nature of the gamma-ray Active Galactic Nuclei of Uncertain type}
\author{F. Massaro\altaffilmark{1}, R. D'Abrusco\altaffilmark{2}, G. Tosti\altaffilmark{3,4}, 
M. Ajello\altaffilmark{1}, D. Gasparrini\altaffilmark{5}, J. E. Grindlay\altaffilmark{2} \& Howard A. Smith\altaffilmark{2}.}

\affil{SLAC National Laboratory and Kavli Institute for Particle Astrophysics and Cosmology, 2575 
	Sand Hill Road, Menlo Park, CA 94025}
\affil{Harvard - Smithsonian Astrophysical Observatory, 60 Garden Street, Cambridge, MA 02138}
\affil{Dipartimento di Fisica, Universit\`a degli Studi di Perugia, 06123 Perugia, Italy}
\affil{Istituto Nazionale di Fisica Nucleare, Sezione di Perugia, 06123 Perugia, Italy}
\affil{ASI Science Data Center, ESRIN, I-00044 Frascati, Italy}

\begin{abstract}
Despite the large number of discoveries made recently by \fer,
the origins of the so called unidentified $\gamma$-ray sources remain unknown. 
The large number of these sources suggests that among them there could be a population that significantly contributes to the
isotropic gamma-ray background and is therefore crucial to understand their nature. 
The first step toward a complete comprehension of the unidentified $\gamma$-ray source population is to identify  
those that can be associated with blazars, the most numerous class of extragalactic sources in the $\gamma$-ray sky.
Recently, we discovered that blazars can be recognized and separated from other extragalactic sources using the infrared (IR) \wse\ satellite colors.
The blazar population delineates a remarkable and distinctive region of the IR color-color space, the \wse\ blazar strip.
In particular, the subregion delineated by the $\gamma$-ray emitting blazars 
is even narrower and we named it as the \wse\ Gamma-ray Strip (WGS).
In this paper we parametrize the \gstrp\ on the basis of a single parameter $s$
that we then use to determine if $\gamma$-ray Active Galactic Nuclei of the uncertain type (AGUs)
detected by \fer\ are consistent with the \gstrp\ and so can be considered blazar candidates.
We find that 54 AGUs out of a set 60 analyzed have IR colors consistent with the \gstrp; only 6 AGUs are outliers.
This result implies that a very high percentage (i.e., in this sample about 90\%) of the AGUs detected by \fer\ are indeed blazar candidates. 
\end{abstract}


\section{Introduction}
\label{sec:intro}
With the recent advent of the \fer\ mission, the $\gamma$-ray astronomy is living a new golden age
with several striking discoveries already performed during the first three years.

According to the second \fer\ $\gamma$-ray LAT catalog \citep[2FGL,][]{abdo11}, \fer\  
detected 1873 sources, 576 of which are still unidentified even if  
the localization of the $\gamma$-ray sources has significantly improved with respect to the past $\gamma$-ray missions.
For this reason, despite the large number of new discoveries already achieved,
the nature of the unidentified $\gamma$-ray sources is still an open question.
This unsolved issue is extremely relevant for the origin of the isotropic gamma-ray background,
since, given the large number of unidentified $\gamma$-ray sources, 
new classes of unknown extragalactic $\gamma$-ray sources that can 
significantly contribute to the isotropic gamma-ray background could be hidden.

On the other hand, the most detected $\gamma$-ray sources in the MeV-GeV energy range belong to 
the rarest class of active galactic nuclei, the blazars.
They are an intriguing class of active galactic nuclei, characterized by non-thermal radiation emitted over the entire electromagnetic spectrum,
and interpreted as arising from a relativistic jet closely aligned to the line of sight \citep[see e.g.][]{blandford78}.
Blazars come in two flavors: the BL Lac objects and the flat spectrum radio quasars, 
where the common discriminating criterion between the two classes is the
equivalent width of the optical emission lines, traditionally
weaker than 5${\AA}$ in the former rather than in the latter \citep{stickel91,stoke91}.
In the following, we indicate the BL Lacs as BZBs and the 
Flat Spectrum radio quasars as BZQs, according to the ROMA-BZCAT nomenclature \citep{massaro09,massaro10}.

\begin{figure}[]
\includegraphics[height=6.4cm,width=8.2cm,angle=-0]{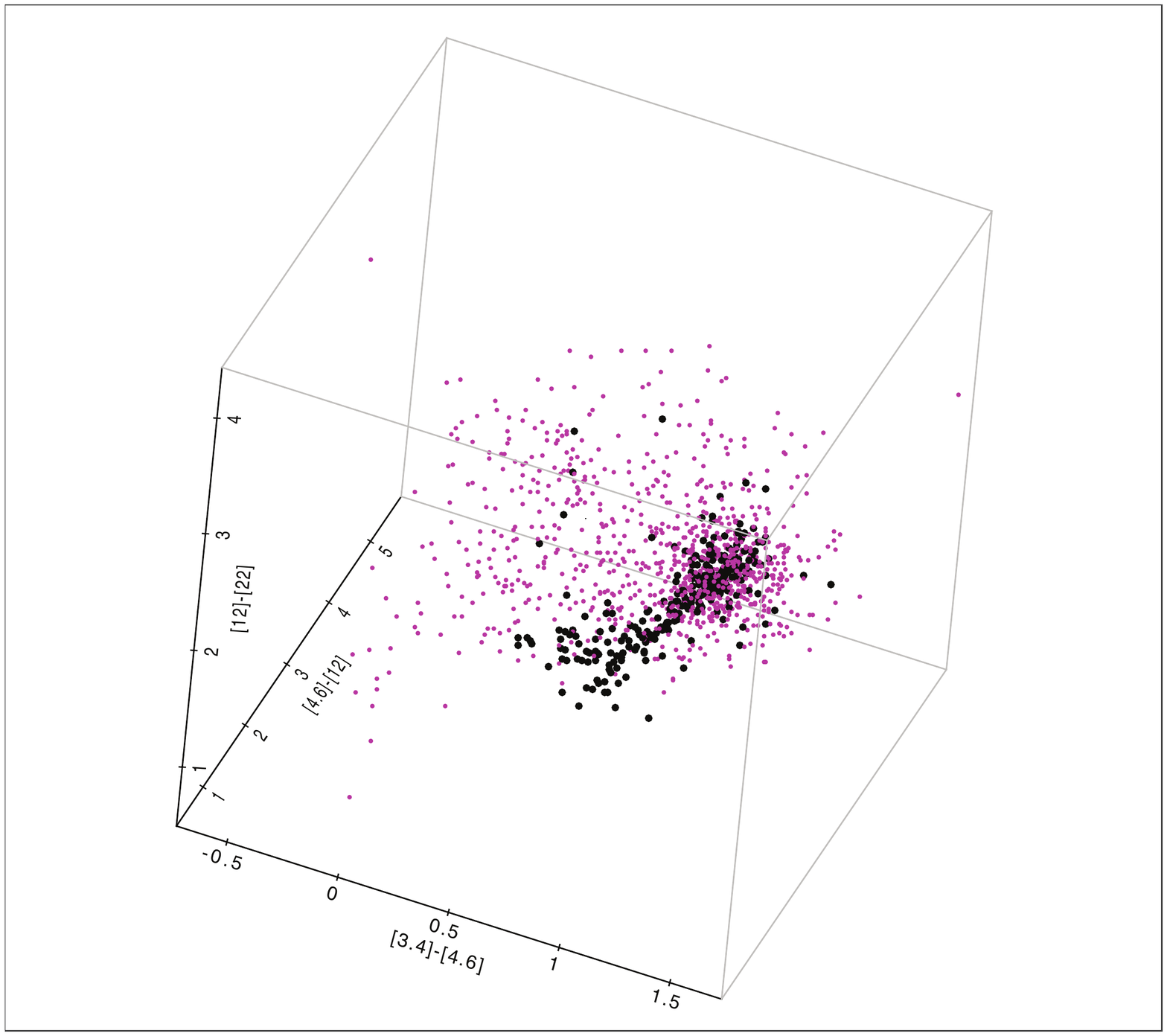}
\caption{The 3D representation of the \wse\ Blazar Strip (blazars are indicated in magenta)
and the subregion of the \gstrp\ ($\gamma$-ray emitting blazars are indicated in black) in the IR colors  
built with the magnitudes in the \wse\ bands at [3.4]-[4.6]-[12]-[22] $\mu$m.}
\label{fig:3D}
\end{figure}

\begin{figure}[b]
\includegraphics[height=6.4cm,width=8.5cm,angle=0]{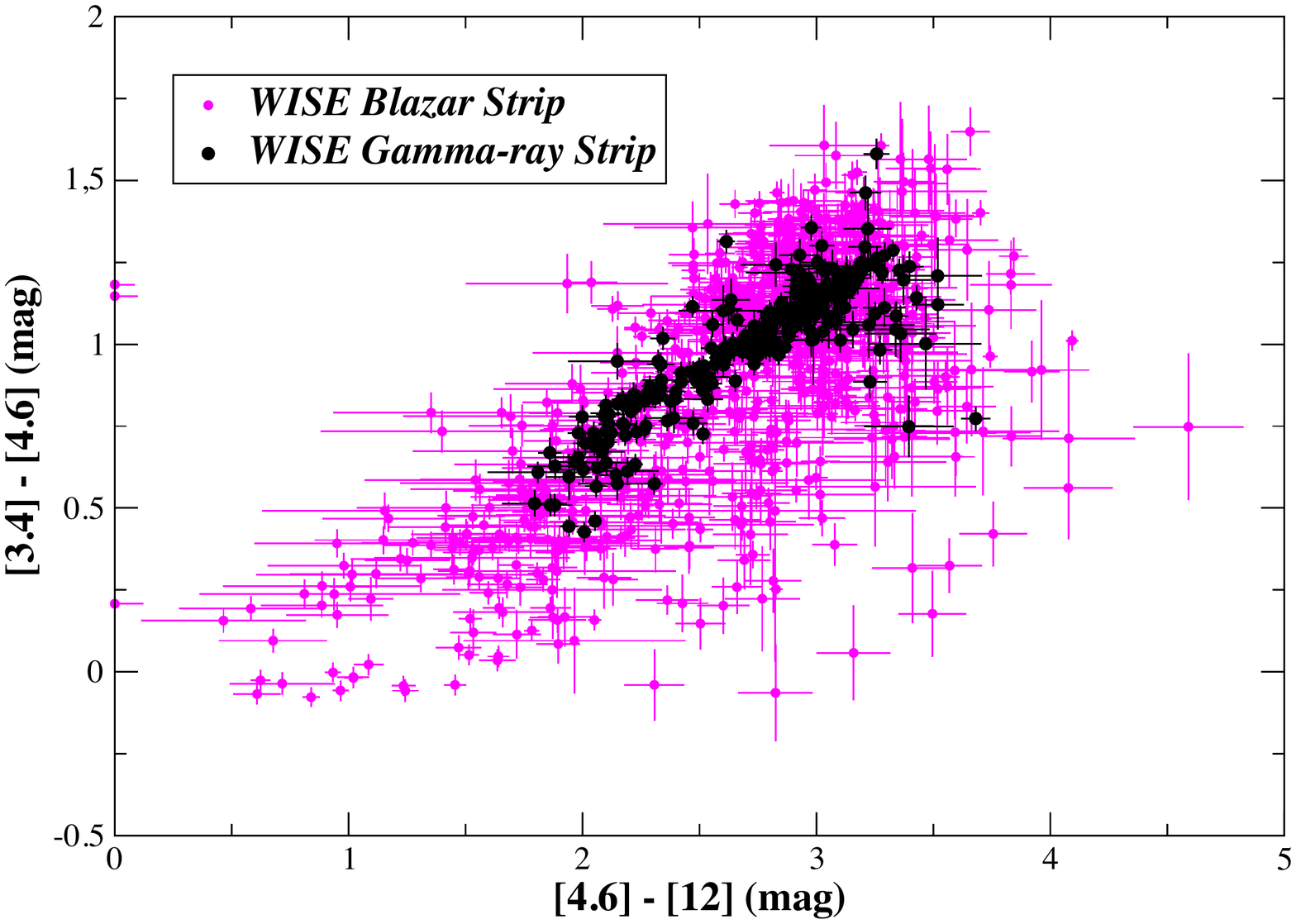}
\caption{The 2D projection of the \wse\ Blazar Strip (blazars are in magenta)
and the subregion of the \gstrp\ ($\gamma$-ray emitting blazars are in black) in the IR color diagram [3.4]-[4.6]-[12] $\mu$m.}
\label{fig:2D}
\end{figure}

Recently, using the preliminary data release of the \wse\ infrared (IR) survey \citep{wright10},
we discovered that IR color-color diagrams 
allows us to distinguish between extragalactic sources dominated 
by non-thermal emission, like blazars, and other classes of galaxies and/or active galactic nuclei
(Massaro et al. 2011, hereinafter Paper I, see also Plotkin et al. 2011).
In particular, the blazar population delineates a tight, distinct region of the IR color space, 
indicated as the \wse\ Blazar Strip (Paper I).
The \wse\ Blazar Strip is a region in the 3D infrared color space delineated by the blazar population.
This region is narrower when considering only the IR colors of the blazars 
that are detected in the $\gamma$-rays, indicated as the \wse\ gamma-ray strip \citep[\gstrp, see][hereinafter Paper II]{dabrusco12}. 
A 3D scatter plot of the \wse\ Blazar Strip and the subregion of the \gstrp\ are shown in the IR diagram
of Figure~\ref{fig:3D}, while the [3.4]-[4.6]-[12] $\mu$m 2D projection is reported in Figure~\ref{fig:2D}.

One of the major difficulties of the association procedures for the \fer\ $\gamma$-ray sources with
active galactic nuclei is that, due to the lack of radio and X-ray informations
and to the large uncertainty on the $\gamma$-ray position, it is not always possible to recognize if
there is a blazar candidate within the positional error region.
Thus, the main aim of this paper is to build a parametrization of the \gstrp\ in order to 
verify whenever $\gamma$-ray sources have been associated to a counterpart that is a blazar candidate,
being consistent with the \gstrp.
In particular, we studied the \wse\ counterparts of the Active galactic nuclei of uncertain type (AGUs),
defined according to the 2FGL and the 2LAC criteria \citep{ackermann11}, and their consistency with the \gstrp.
The AGUs are defined as the radio and/or X-ray counterparts of $\gamma$-ray sources 
associated to by the 2FGL Likelihood Ratio method, but without a good optical spectrum that enable their classification
\citep{abdo11}.
This paper is organized as follows.
Section~\ref{sec:parametrization} describes the procedure adopted to parametrize the \gstrp.
Section~\ref{sec:counterpart} we discuss on the consistency of the \wse\ counterparts of the AGU sample with the \gstrp
while Section~\ref{sec:kde} describes the non-parametric analysis of the \gstrp\ based on the Kernel Density Estimation (KDE).
In Section~\ref{sec:effects} we investigated possible selection effects that could affect our \gstrp\ parametrization.
Our conclusions are discussed in Section~\ref{sec:summary}.

\section{The \wse\ gamma-ray strip parametrization}
\label{sec:parametrization}

\subsection{The sample selection}
\label{sec:sample}
We use the sample of $\gamma$-ray emitting blazars already selected in Paper II to parametrize the \gstrp.

This sample was selected from the 2FGL, that contains 805 sources associated with a blazar: 
435 BZBs and 370 BZQs respectively.
However only 659 (347 BZBs and 312 BZQs) of these 
are listed and classified according to the criteria used in the ROMA-BZCAT \citep[e.g.,][]{massaro09}.
We excluded from our analysis all the blazars with a \fer\ analysis flag, 
according to the 2FGL and the 2LAC \citep{abdo11,ackermann11}.

In particular, 329 (164 BZBs and 165 BZQs) blazars, out of the original 659,
lie in the portion of the sky reported in the \wse\ preliminary source catalog,
but only 296 (143 BZBs and 153 BZQs) have a \wse\ counterpart
within 2.4$^{\prime\prime}$ radius (see Paper I).
To be more conservative, we excluded from our analysis 12 blazars 
(8 BZBs and 4 BZQs) with respect to the 296 blazars in sample selected in Paper II, 
because they have a 95\% upper limit on the \wse\ magnitude at 22 $\mu$m.
Then, we use this 2FB sample composed of the 284 blazars (135 BZBs and 149 BZQs) to build the \gstrp\ parametrization.
We notice that all the selected blazars also belong the 2LAC sample \citep{ackermann11}.
 
According to the classification available in ROMA-BZCAT the blazars of uncertain type
have been excluded from our analysis, while the BL Lac candidates have been considered as BZBs.
More details about the 2FB sample and the source selections are given in Paper II.

Finally, we emphasize that our selection is based only on $\gamma$-ray blazars that belong to the ROMA-BZCAT 
because this is the largest catalog of blazars available in literature in which each source 
is spectroscopically classified at optical frequencies.

\subsection{The \wse\ blazar associations}
\label{sec:associations}
The IR color-color diagrams have been built using the archival \wse\ Preliminary Source Catalog, 
that covers $\sim$ 57\% of the sky
\footnote{wise2.ipac.caltech.edu/docs/release/prelim/preview.html} .
The \wse\ mission mapped the sky at 3.4, 4.6, 12, and 22 $\mu$m 
in 2010 with an angular resolution of 6.1, 6.4, 6.5 \& 12.0$^{\prime\prime}$ in the four bands, achieving 5$\sigma$ 
point source sensitivities of 0.08, 0.11, 1 and 6 mJy respectively in unconfused regions on the ecliptic. 
All the \wse\ magnitudes are in the Vega system.
In particular, the absolute (radial) differences between \wse\ source-peaks and ``true" astrometric positions 
anywhere on the sky are no larger than $\sim$ 0.50, 0.26, 0.26, and 1.4$^{\prime\prime}$ in the
four \wse\ bands, respectively \citep{cutri11}\footnote{wise2.ipac.caltech.edu/docs/release/prelim/expsup/sec2\_3g.html}.

For our analysis, unless stated otherwise, we considered only \wse\ sources detected 
with a minimum signal-to-noise ratio of 7 in at least one band
\footnote{We take the opportunity to correct here an error that appears in Paper II. 
The sources of the 2FB sample are detected with a minimum signal-to-noise ratio of 7 in at least one band 
rather than in all four bands as reported in Paper II.}.
The positional coincidences of blazars in the observed \wse\ sky have been searched 
within a circular region of radius 2.4$^{\prime\prime}$.
This corresponds to the combination of the error of 1$^{\prime\prime}$, 
assumed for the radio position reported in the ROMA-BZCAT \citep{massaro09} and taking into account of 
astrometric uncertainties in the \wse\ preliminary data,
and the positional error of the fourth \wse\ band at 22$\mu$m (i.e., 1.4$^{\prime\prime}$) (see also Paper I).
All the associations of the 2FB blazars with \wse\ sources 
are unique and no multiple matches have been found (see Papers I and II for more details).
The chance probabilities of the \wse\ associations for the sources in the 2FGL and in the ROMA-BZCAT are reported in Paper II.

\subsection{The \wse\ gamma-ray strip projections}
\label{sec:projection}
We built the parametrization of the \gstrp\ considering only the sources in the 2FB sample 
(see Section~\ref{sec:sample}) and using the three different 2D projections of the \gstrp\ 
delineated in the [3.4]-[4.6]-[12] $\mu$m, [4.6]-[12]-[22] $\mu$m, [3.4]-[4.6]-[22] $\mu$m
color-color planes.
In each color-color 2D projection we determined the smaller irregular quadrilateral  
containing at least 95\% of the blazars in the 2FB sample considering their position within error 
(see also Section~\ref{sec:parameter} for more details).
The irregular quadrilateral defining the \gstrp\ subregions have been drawn by hand.
The KDE analysis has been used to verify, {\it a posteriori}, 
that the hand-drawn boundaries of the \gstrp\ are in agreement with the sharp decline in density of \gstrp\ sources, 
as evaluated by this non parametric method {\bf (see Section~\ref{sec:kde} for more details)}.
This \gstrp\ modeling has been developed separately for the BZBs and the BZQs, and
for all their three 2D projections.

\begin{figure}[]
\includegraphics[height=6.2cm,width=8.5cm,angle=0]{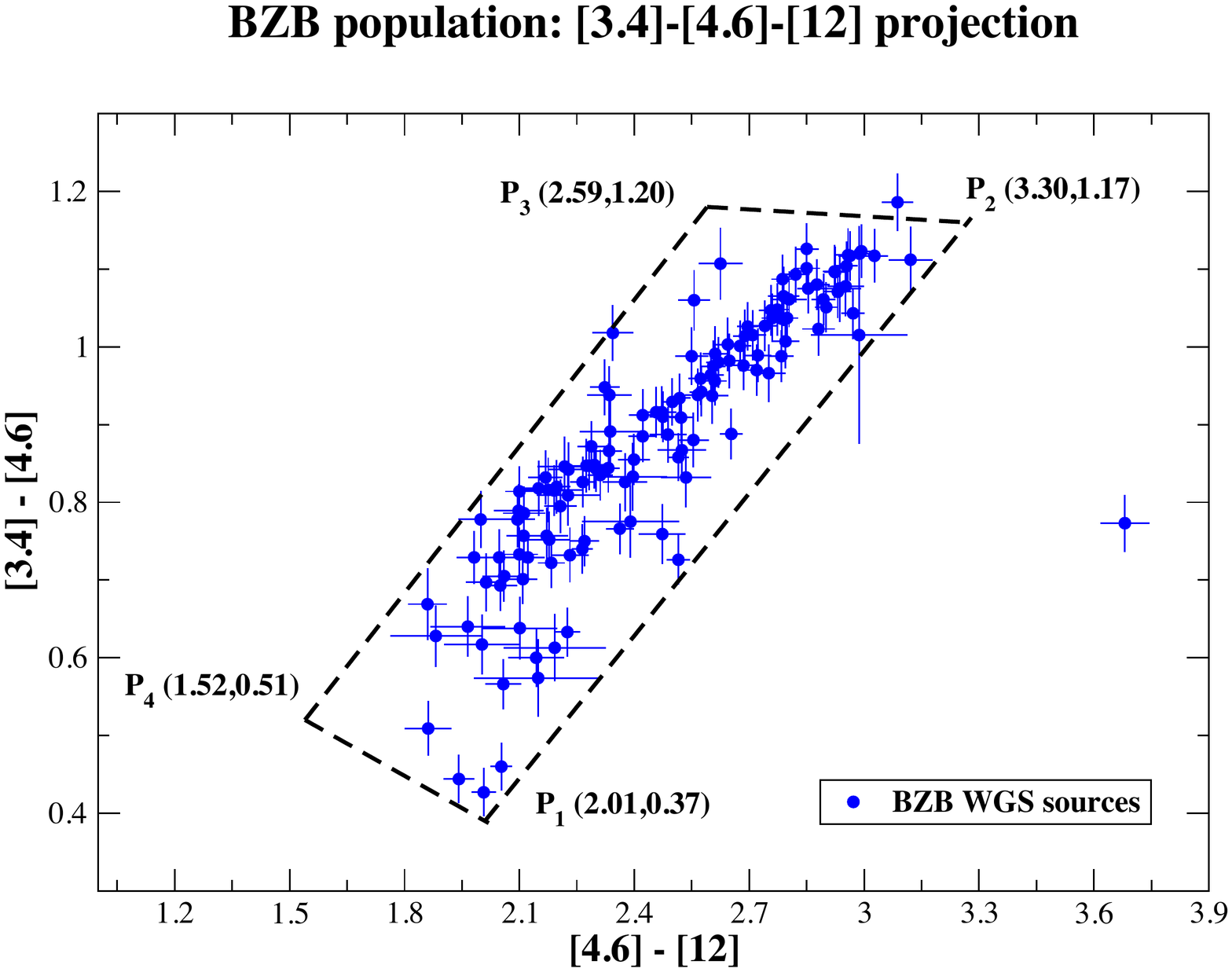}
\caption{The [3.4]-[4.6]-[12] $\mu$m 2D projection of the \gstrp\ in the subregion of the BZB population is shown.}
\label{fig:strip_bzb_pln1}
\end{figure}

\begin{figure}[]
\includegraphics[height=6.2cm,width=8.5cm,angle=0]{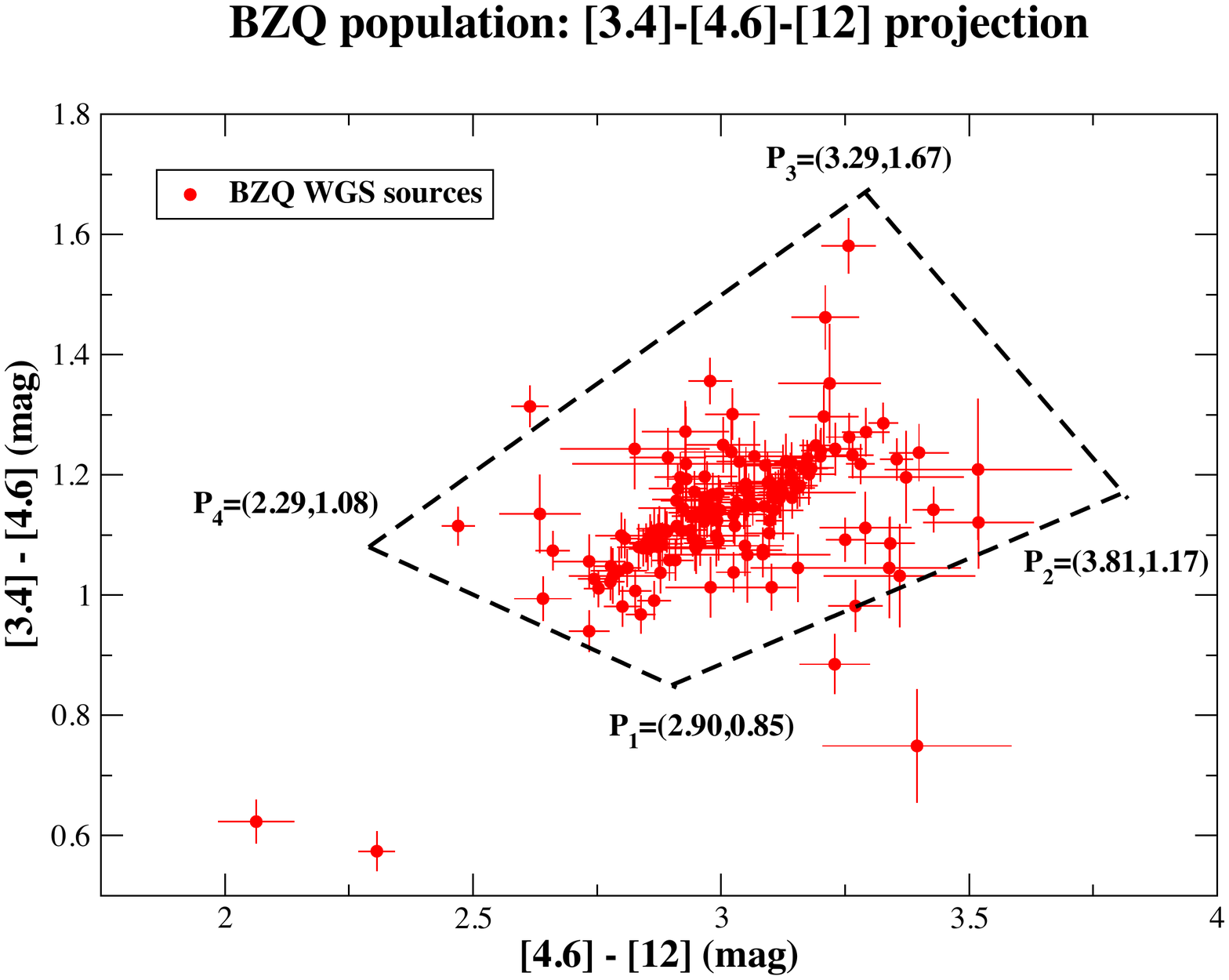}
\caption{Same of Figure~\ref{fig:strip_bzq_pln1} but for the case of the BZQ population.}
\label{fig:strip_bzq_pln1}
\end{figure}

\begin{figure}[]
\includegraphics[height=5.6cm,width=8.5cm,angle=0]{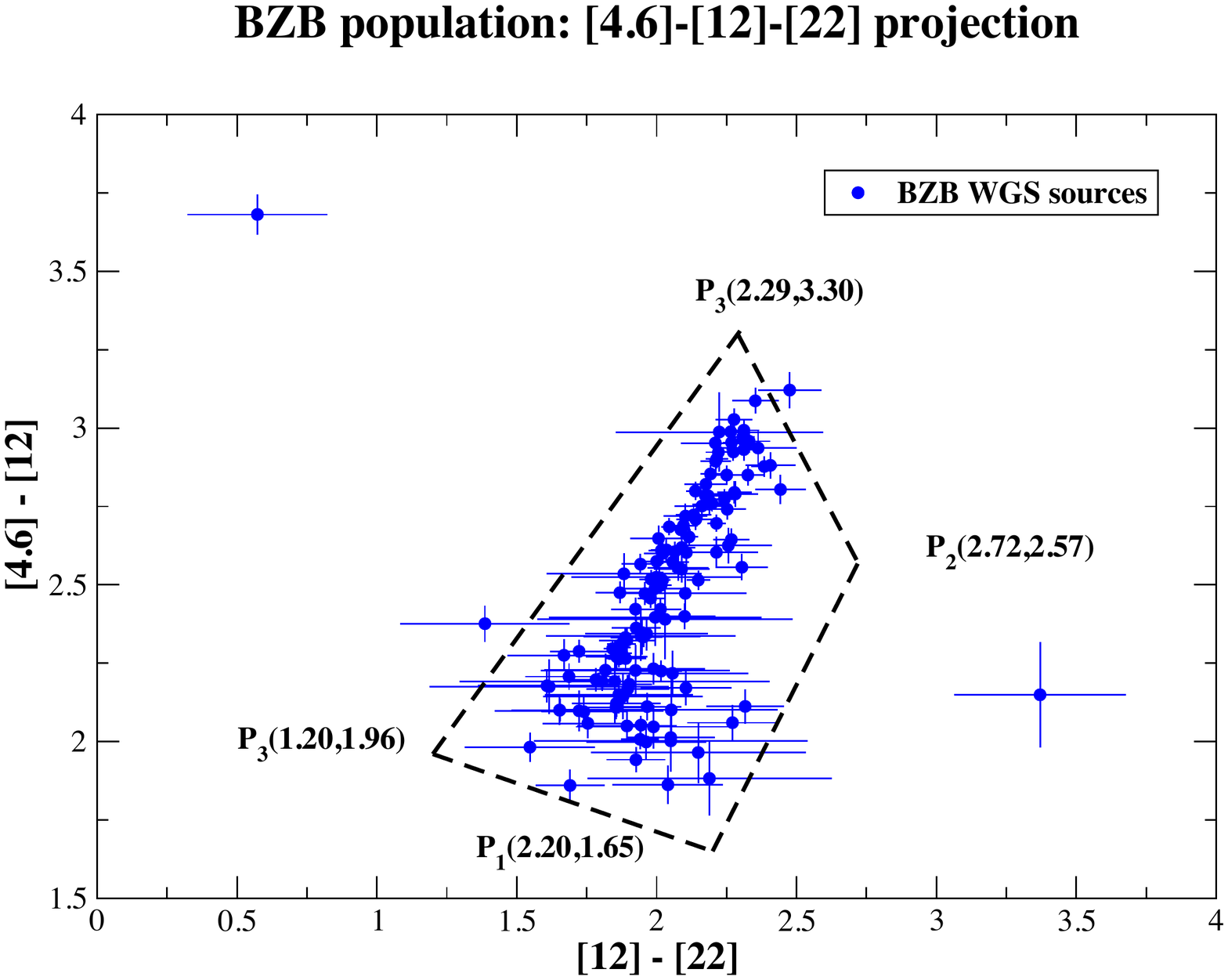}
\includegraphics[height=5.6cm,width=8.5cm,angle=0]{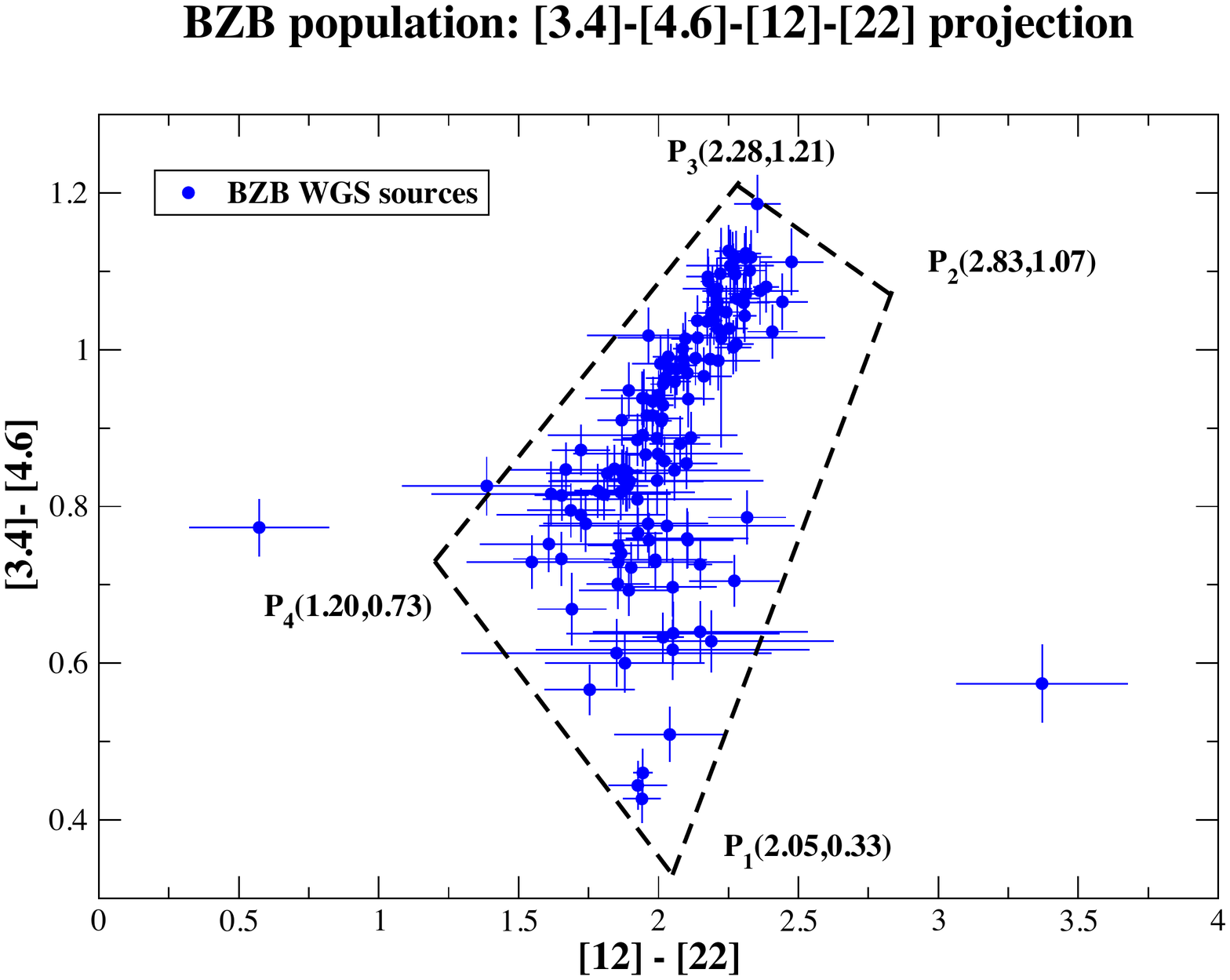}
\caption{Same of Figure~\ref{fig:strip_bzb_pln1} for the case of the BZB population on the \gstrp\
in the two remaining different color-color projections.}
\label{fig:strip_bzb_pln2-3}
\end{figure}

\begin{figure}[]
\includegraphics[height=5.6cm,width=8.5cm,angle=0]{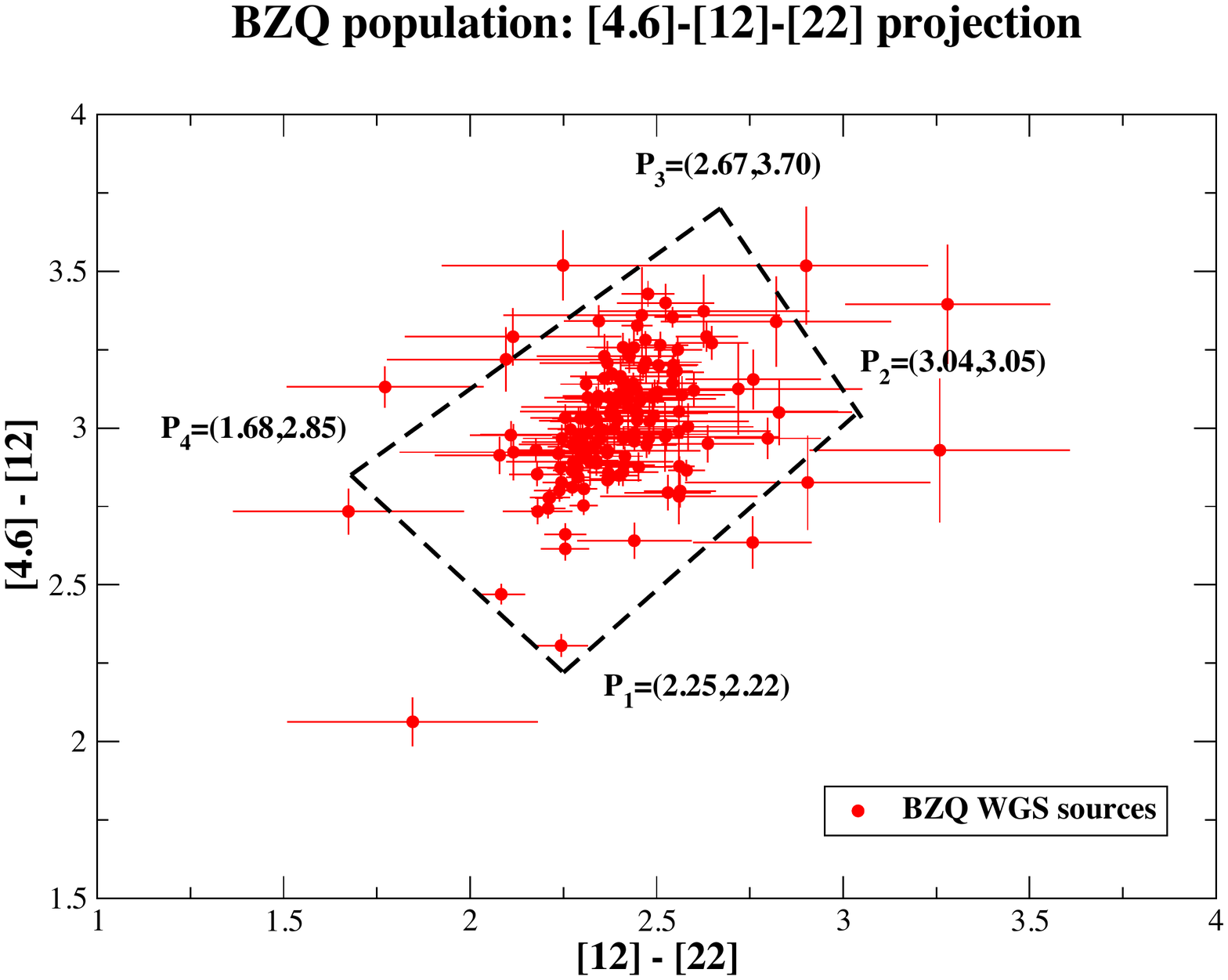}
\includegraphics[height=5.6cm,width=8.5cm,angle=0]{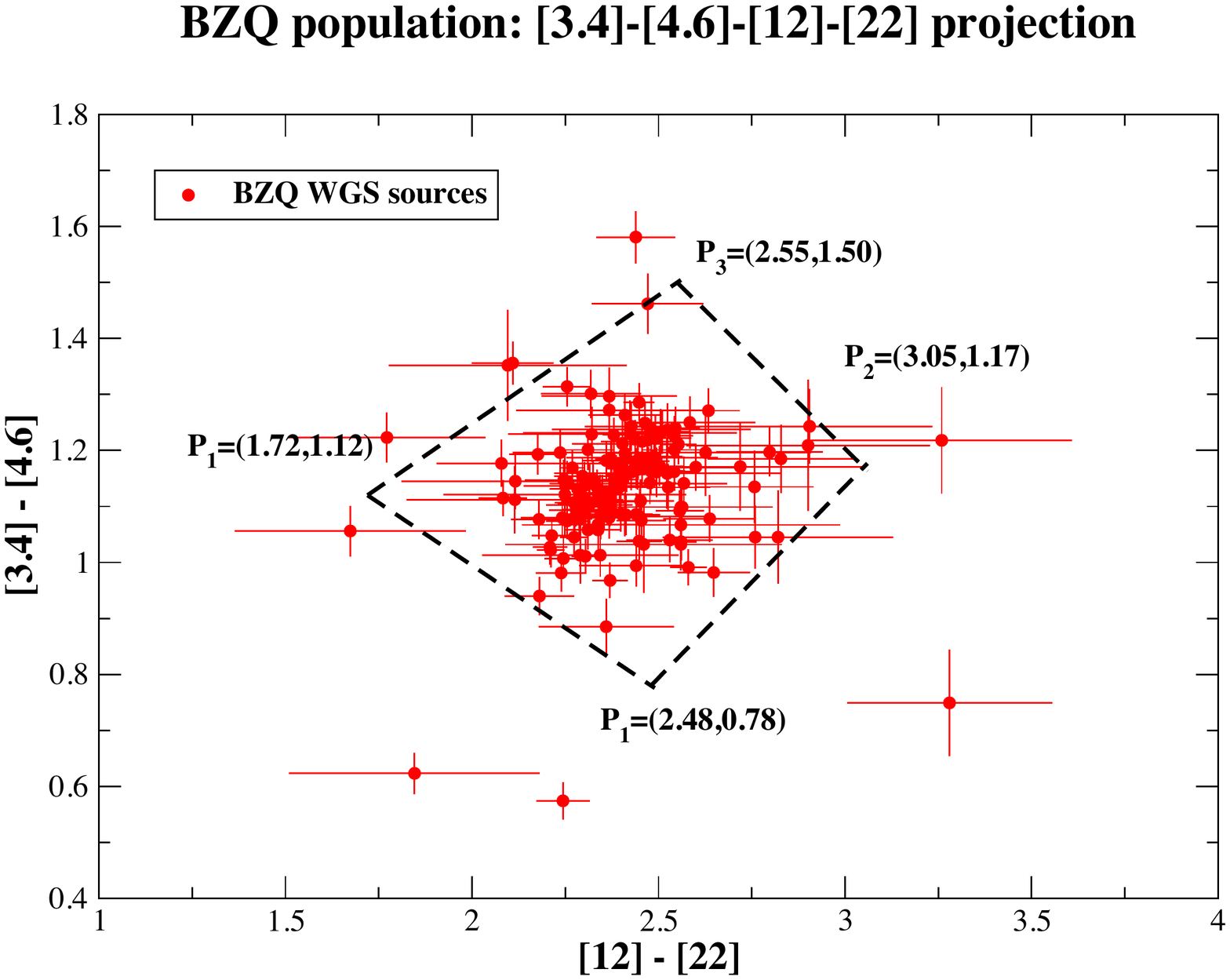}
\caption{Same of Figure~\ref{fig:strip_bzq_pln1} for the case of the BZQ population on the \gstrp\
in the two remaining different color-color projections.}
\label{fig:strip_bzq_pln2-3}
\end{figure}

The [3.4]-[4.6]-[12] $\mu$m, [4.6]-[12]-[22] $\mu$m and [3.4]-[4.6]-[12]-[22] $\mu$m 2D projections 
of the \gstrp\ for the BZB and the BZQ populations are shown in Figure~\ref{fig:strip_bzb_pln1}, Figure~\ref{fig:strip_bzq_pln1}, 
Figure~\ref{fig:strip_bzb_pln2-3} and Figure~\ref{fig:strip_bzq_pln2-3}), respectively.

In the following we also report the boundaries chosen for our \gstrp\ parametrization.
For the BZB projections the extremal points of the \gstrp\ have coordinates:
$P_1$=(2.01,0.37), $P_2$=(3.30,1.17), $P_3$=(2.59,1.20), $P_4$=(1.52,0.51)
in Figure~\ref{fig:strip_bzb_pln1},
$P_1$=(2.20,1.65), $P_2$=(2.72,2.57), $P_3$=(2.29,3.30), $P_4$=(1.20,1.96),
in Figure~\ref{fig:strip_bzb_pln2-3} (upper panel), while
$P_1$=(2.05,0.33), $P_2$=(2.83,1.07), $P_3$=(2.28,1.21), $P_4$=(1.20,0.73),
in Figure~\ref{fig:strip_bzb_pln2-3} (lower panel).

On the other hand, for the BZQ projections the extremal points of the \gstrp\ have coordinates:
$P_1$=(2.90,0.85), $P_2$=(3.81,1.17), $P_3$=(3.29,1.67), $P_4$=(2.29,1.08)
in Figure~\ref{fig:strip_bzq_pln1},
$P_1$=(2.25,2.22), $P_2$=(3.04,3.05), $P_3$=(2.67,3.70), $P_4$=(1.68,2.85),
in Figure~\ref{fig:strip_bzq_pln2-3} (upper panel), while
$P_1$=(2.48,0.78), $P_2$=(3.05,1.17), $P_3$=(2.55,1.50), $P_4$=(1.72,1.12),
in Figure~\ref{fig:strip_bzq_pln2-3} (lower panel).

\subsection{The strip parameter $s$}
\label{sec:parameter}
\begin{figure}[]
\includegraphics[height=6.7cm,width=8.5cm,angle=0]{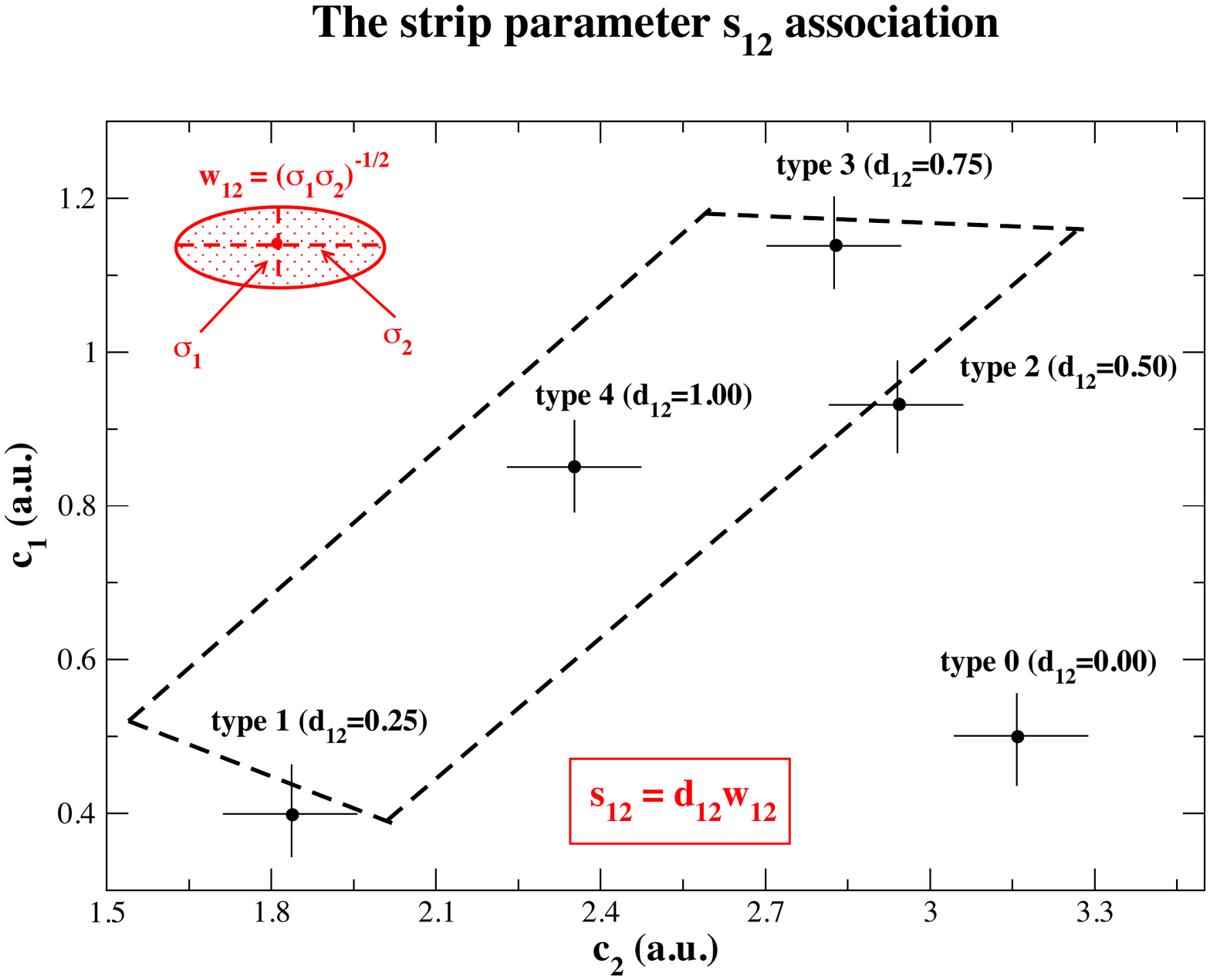}
\caption{The schematic view of the strip parametrization in the example of the 
$c_{1}$-$c_{2}$ 2D projection in arbitrary units (a.u.). 
We report the method described in Section~\ref{sec:parameter}
to assign to each point of the 2D projection of the \gstrp\ a value of the {\it discrete strip parameter} $d_{12}$ 
and the associated value of the {\it weight strip parameter} $w_{12}$.
The combination of these two values provides the {\it continuos strip parameter} $s_{12}$ 
for each given source (see Equation~\ref{eq:main}).}
\label{fig:scheme}
\end{figure}

To illustrate the \gstrp\ parametrization we consider the schematic case of the first projection:
[3.4]-[4.6], [4.6]-[12], hereinafter $c_{1}$-$c_{2}$ with the correspondent errors
$\sigma_{1}$ and $\sigma_{2}$, respectively (see Figure~\ref{fig:scheme}).

Based on the \wse\ source location in the $c_{1}$-$c_{2}$ diagram we can distinguish 5 types of objects.
Each source, given its IR colors, corresponds to a single point in each 2D color-color projection of the \gstrp.
However, including the errors on both axes, it is represented by a cross with 4 {\it extremal points},
calculated considering the $\pm$1$\sigma$ error on each color.
Then, we can define five different type of sources, 
according to the schematic view shown in Figure~\ref{fig:scheme}: 
\begin{itemize}
\item{{\it type 4}: sources with all the extremal points within the \gstrp\ projection;}
\item{{\it type 3}: sources for which only 3 extremal points lie within the region of the \gstrp;} 
\item{{\it type 2}: sources with only two extremal points consistent with the \gstrp;}
\item{{\it type 1}: sources with only a single extremal point associated with the \gstrp;}
\item{{\it type 0}: sources without extremal points of the error cross on the \gstrp.}
\end{itemize}
We can assign to each type of source a {\it discrete strip parameter} $d$
ranging between 0 and 1, according to the scheme illustrated in Figure~\ref{fig:scheme}.
For example, in the case of the $c_{1}$-$c_{2}$ projection of the \gstrp,
we assign to type 4 sources a value $d_{12}$=1, while source of type 0 corresponds to $d_{12}$=0. 
For the same 2D projection, the intermediate values have been assigned as follows:
type 3 have $d_{12}$=0.75, type 2 have $d_{12}$=0.5 and type 1 $d_{12}$=0.25.

On the same $c_{1}$-$c_{2}$ diagram,
we also assign a {\it weight strip parameter} $w_{12}$ to each value of the $d_{12}$ parameter
defined as:
$w_{12}$= $(\sigma_{1}\,\sigma_{2})^{-1/2}$, proportional to the area of the ellipse described 
by the error bars of each point (see inset of Figure~\ref{fig:scheme} for more details).
Then, we define the {\it continuos strip parameter} $s_{12}$ as:
\begin{equation}
s_{12} = d_{12}\,w_{12}.
\label{eq:main}
\end{equation}
We note that the parameter $w_{12}$ has been chosen to take into account the different errors on both axes when 
comparing two sources that might belong to the same type.
It also allow us to make the $s_{12}$ continuos rather than discrete as $d_{12}$.

We repeated the entire procedure described above for each
2D projection of the \gstrp: $c_{1}$-$c_{2}$, $c_{2}$-$c_{3}$ and $c_{1}$-$c_{3}$,
generating the values of the continuos strip parameters $s_{12}$, $s_{23}$, $s_{13}$, respectively. 
Then, all these values of the strip parameters for three different 2D projections have been combined together to define
an unique {\it total strip parameter} $s$. 
The {\it total strip parameter} is the geometric average of the $s$ values of each 2D projection:
\begin{equation}
s = (s_{12}\,s_{23}\,s_{13})^{1/3}\, .
\end{equation}
We emphasize that sources that lie outside of the \gstrp\ in at least one of its 2D projection have one of the correspondent 
$s_{12}, s_{23}, s_{13}$ parameter equal to zero and consequently the total $s$ value is null as well.
This occurs because the discrete $d$ parameter is zero for sources outside the \gstrp\ (see Figure~\ref{fig:scheme}).

We divided all the $s$ parameters for the maximum $s$ values of the BZBs and BZQs that lie on the \gstrp\
to re-normalize $s$ in the range 0 and 1.
This re-normalization can be applied to the $s$ values of all the \wse\ sources, because those outside the \gstrp\
will have $s$ null by definition.

The $s$ parameter represents an estimate of the {\it distance}, 
in the IR colors parameter space and weighted with the errors on each axes,
between the \gstrp\ and a generic \wse\ source, that could potentially belong to it;
this $s$ parameter is different from zero only in the case 
in which the all color error bars of a \wse\ source are consistent with the \gstrp. 
Therefore these $s$ values can be used to rank each IR \wse\ source according
to their {\it association} to the \gstrp.

Finally, we note that to test if a generic \wse\ source has IR colors consistent with the BZBs or with the BZQs 
subregion of the \gstrp, the total strip parameters are indicated as $s_b$ and $s_q$, respectively.
We introduced the above divisions for the $s$ parameters because in future works this allow us to verify
if a generic \wse\ source is more consistent to be a BZB or a BZQ, enabling a classification for new
IR sources that could lie on the \gstrp being $\gamma$-ray blazar candidates.

\subsection{The $s$ parameter distributions}
\label{sec:distribution}
We considered a sample composed of all the \wse\ sources lying in two circular regions of 1 deg radius,
centered at high and low Galactic latitude $b$,  
with the center coordinates of $(l,b)$=(255,-55) deg and $(l,b)$=(338,-1) deg, respectively.
These sources do not have upper limits on their \wse\ magnitude values and are detected with a signal to noise ratio $>$ 7
in at least one band as for the blazars in the 2FB sample.

\begin{figure}[]
\includegraphics[height=5.6cm,width=8.5cm,angle=0]{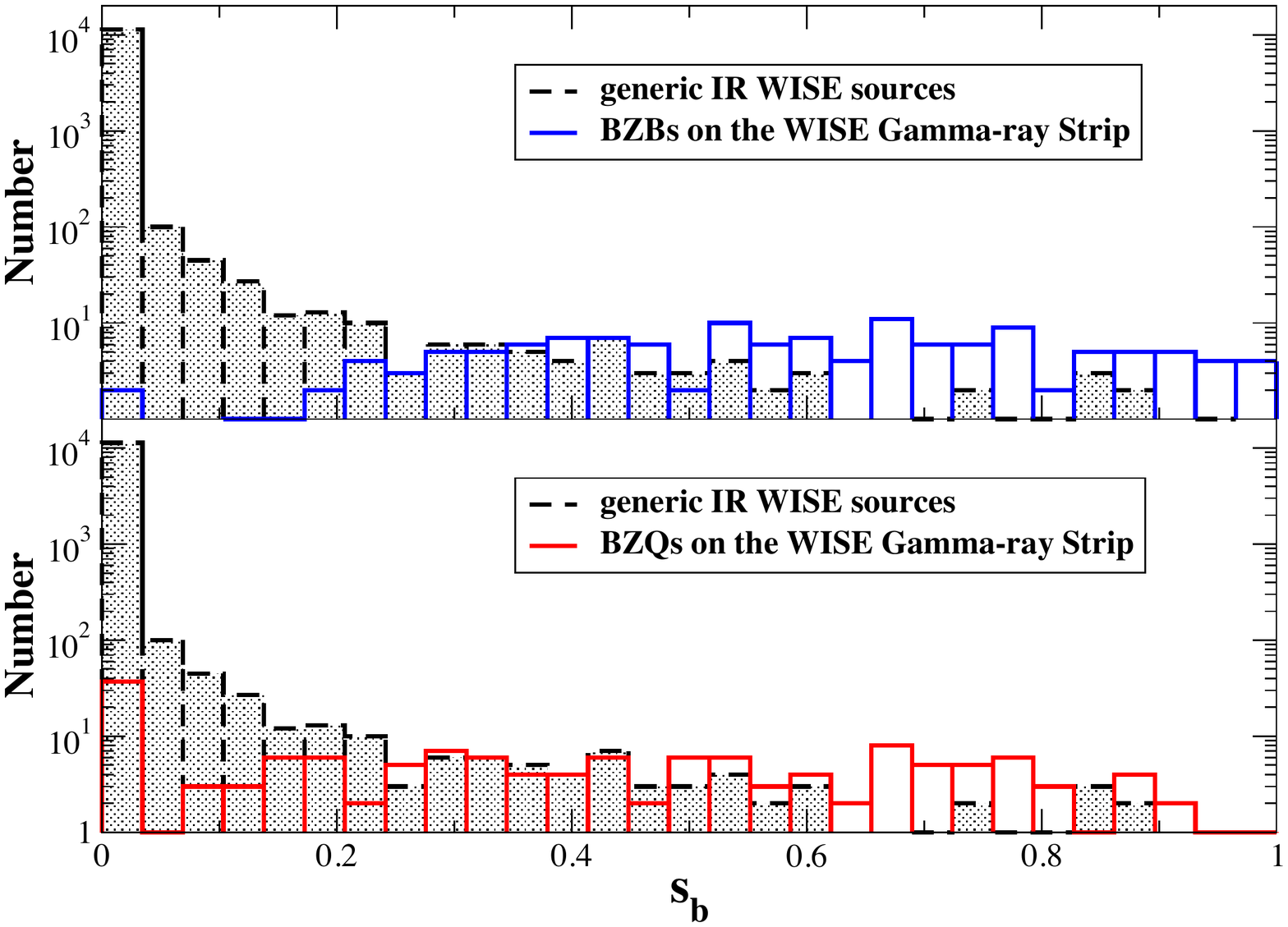}
\caption{The distribution of the strip parameter $s_b$ for the BZBs (blue) and the BZQs (red)
that lie on the \gstrp\ in comparison with the generic IR \wse\ sources (black).}
\label{fig:bzb_wise}
\end{figure}

\begin{figure}[]
\includegraphics[height=5.3cm,width=8.5cm,angle=0]{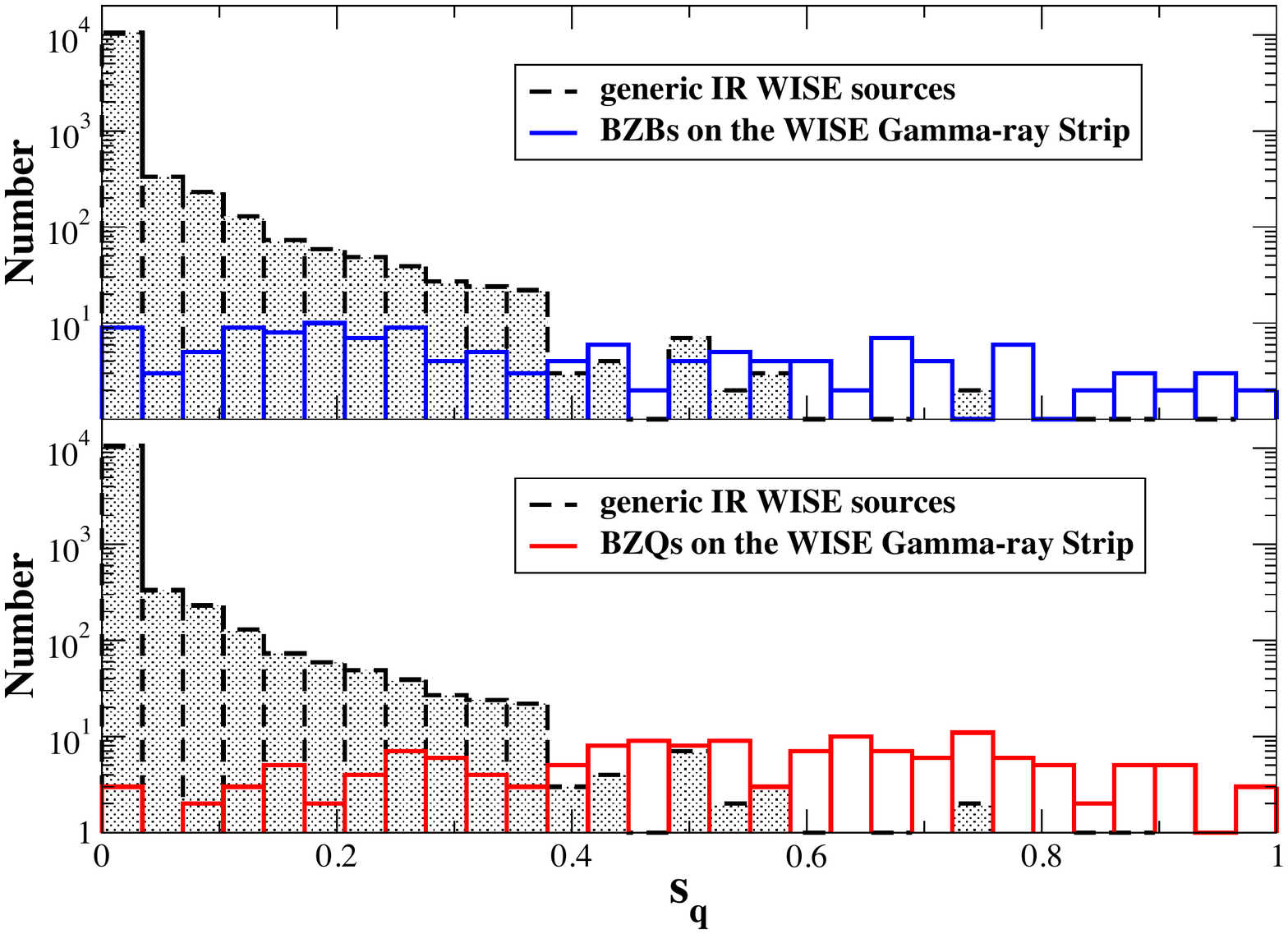}
\caption{Same of Figure~\ref{fig:bzb_wise} for the distribution of the $s_q$ parameter.}
\label{fig:bzq_wise}
\end{figure}

We calculated the $s$ parameters for all the 11599 generic IR \wse\ sources.
This analysis provides an estimate of the probability to find a generic \wse\ source 
in the sky with a particular value of $s_b$ and/or $s_q$.
We perform this analysis considering the distinction between the two blazar classes (i.e., BZBs and BZQs),
The distributions of the $s_b$ and $s_q$ 
parameters for the BZBs and the BZQs that lie on the \gstrp\, in comparison with the generic IR \wse\
sources are shown in Figure~\ref{fig:bzb_wise} and in Figure \ref{fig:bzq_wise}, respectively.

From the distributions of the $s_b$ and $s_q$ parameters for the generic IR \wse\ sources,
we note that 99.9\% of them have $s_b<$0.24 and $s_q<$0.38.
Then, for the BZBs in the 2FB sample only 6 sources out of 135 have $s_b<$ 0.24, and in the case of the BZQs
only 33 sources out of 149 show $s_q$ values lower than 0.38.
We also note that 99.0\% of the generic IR \wse\ sources have $s_b<$0.10 and only 2 BZBs are below this value,
while 97.2\% of the generic IR \wse\ sources together with only 5 BZQs out of 149 have $s_q<$0.14.
Finally, on the basis of the above $s$ distributions we define the {\it outliers} of the \gstrp,
\wse\ sources that have values of $s_b<$0.10 or $s_q<$0.14.

We recognize that the above choice of the $s_b$
and $s_q$ thresholds are extremely conservative.
This choice has been made on the basis of the actual sky coverage of the \wse\ preliminary data release.
At the present status of our analysis we are not able to investigate the IR emission of all the blazars that are
listed in the 2FGL, and the 2FB sample used to build the \gstrp\ parametrization is small with respect to
how it will be available when the \wse\ full archive will be released. 
Consequently, regarding the choice of the threshold values for the $s$ parameters,
we preferred the efficiency to the completeness of our method 
selecting the $s$ limiting values from their distributions in low galactic latitude regions,
even if this choice could increase the possible contamination of the \gstrp.
A deeper investigation of this problem will be considered in future as an {\it a posteriori} analysis of \gstrp\ parametrization.
In particular, when the \wse\ full release will be available, we will improve our method taking into account of the IR source density
at different galactic latitudes and of the varying depth of the exposure for the \wse\ observations.

\section{The AGU counterparts on the \gstrp}
\label{sec:counterpart}
We considered the sample composed of all the AGUs
already classified in the analysis of the 2FGL and the 2LAC \citep{abdo11,ackermann11}. 
According to the 2FGL, the AGUs could be all blazar candidates without a good optical spectrum 
or without an optical spectrum at all \citep{abdo11,ackermann11}. 

We selected the AGUs that lie in the portion of the sky surveyed by \wse\ during the first year, corresponding to 148 \fer\ sources.
Then, we excluded from our analysis all the AGUs with a \fer\ analysis flag, according to the 2FGL and the 2LAC.

The association between each AGU counterpart and the \wse\ sources have been evaluated on the basis of the
same criterion chosen for the blazars on the \wse\ Blazar Strip (see Section~\ref{sec:associations} and Paper I for more details),
considering the position of the radio counterpart for each AGU as reported in the 2FGL and/or in the 2LAC.
There are 60 AGUs out of 148 for which there is a unique association 
with a \wse\ source (see Section~\ref{sec:counterpart})
within the usual region of 2.4$^{\prime\prime}$ radius and with a chance probability of 0.008, 
estimated adopting the method described in Maselli et al. (2010) \citep[see also][and Paper I]{maselli11}
and without upper limits on the \wse\ magnitudes within the \wse\ preliminary data release.

Subsequently, we used the IR colors of the AGU counterparts, as associated in the 2LAC, 
to verify if the \wse\ counterparts of the $\gamma$-ray sources in the 60 AGU sample lie on the \gstrp,
evaluating their $s$ values following the procedure described in Section~\ref{sec:parametrization}.

We found that 6 outliers do not belong to the \gstrp\ out of 60 AGUs,
according to the inclusion based on the threshold values of $s_b<$0.10 or $s_q<$0.14.
With this analysis on the 60 AGU sample, we have been able to check if the association provided
by the 2FGL corresponds to a blazar lying on the \gstrp.

\begin{figure}[]
\includegraphics[height=6.6cm,width=8.5cm,angle=0]{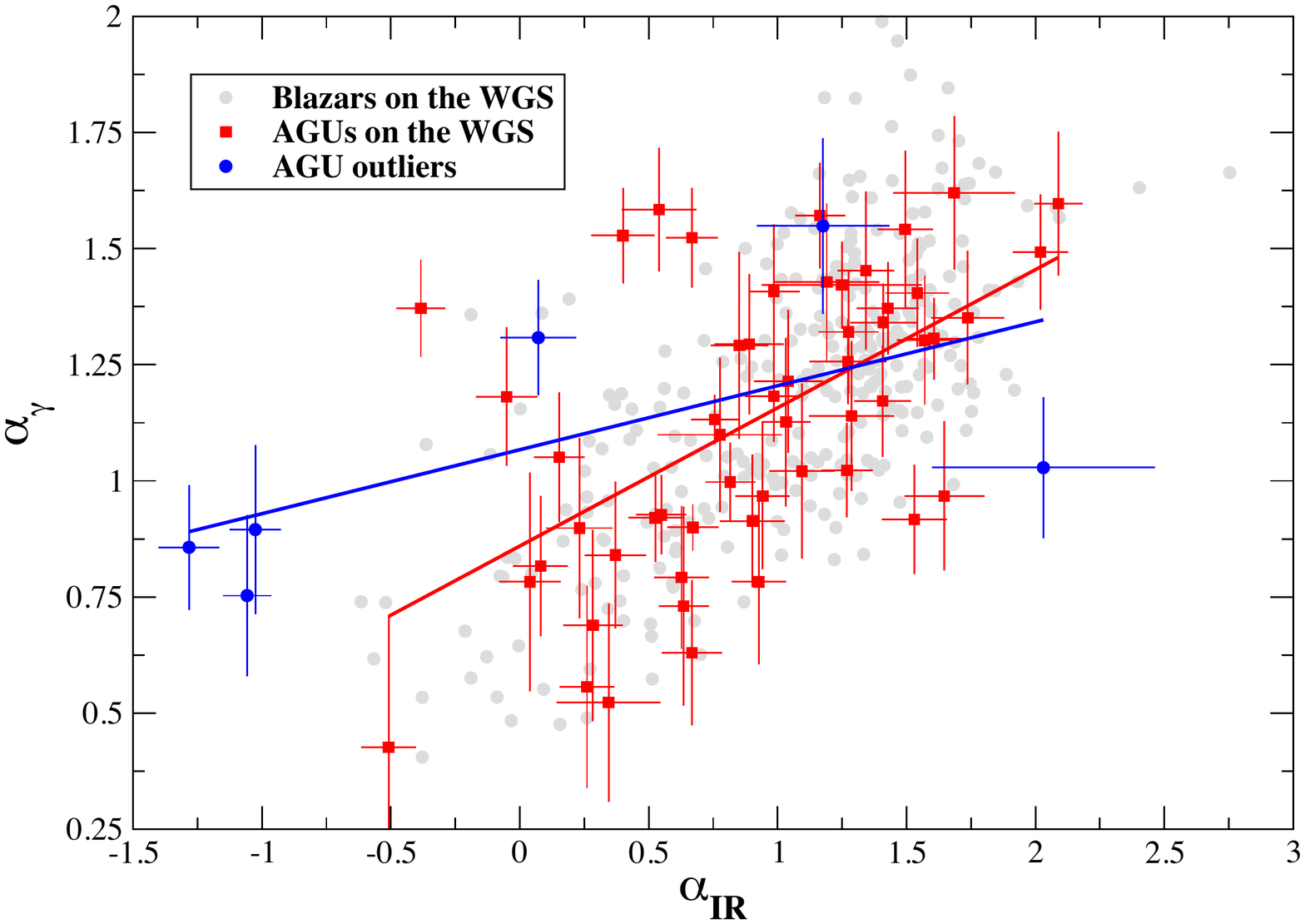}
\caption{The correlation between $\alpha_{IR}$ and $\alpha_{\gamma}$ for the \wse\ counterparts of the AGUs.
The background grey circles represent the correlation found for the \gstrp\ blazars (see Paper II), 
while the red filled squares are the AGUs that have been found consistent with the \gstrp\ accordingly with our parametrization. 
The remaining 6 outliers (blue filled circles) show a weaker and different correlation than the other samples.}
\label{fig:alfas}
\end{figure}

Finally, we estimated the IR spectral index $\alpha_{IR}$ using the [3.4]-[4.6] $\mu$m color
according to Eq. (1) of Paper II,
and we evaluated the correlation between $\alpha_{IR}$ and the spectral index of the associated 2FGL source $\alpha_{\gamma}$.
We found a linear correlation between $\alpha_{IR}$ and $\alpha_{\gamma}$ for the 54 AGU that lie on the \gstrp, with
a correlation coefficient $\rho$= 0.56 and a chance probability of 8.96$\times$10$^{-6}$ and a slope $m$=0.30$\pm$0.06,
that is consistent with that of the \gstrp\ blazars ($\rho$=0.68, $m$=0.36$\pm$0.02, see Paper II) within one sigma (see Figure~\ref{fig:alfas}).
On the other hand, the 6 outliers have a weaker linear correlation between the two spectral indices then the previous sample
with $\rho$=0.40 (chance probability of 0.08) and $m$=0.12$\pm$0.07, different from that of the blazars on the \gstrp\ 
(see Figure~\ref{fig:alfas}).
Finally, in Table~\ref{tab:outliers} and in Table~\ref{tab:candidates} we report the colors, the IR spectral indices 
and the $s$ parameters together with the 2FGL name and the \wse\ and the counterpart names of each AGU analyzed. 
The class of each AGU as derived from the 2LAC analysis is also indicated \citep{ackermann11}.

\section{An independent non-parametric analysis: the kernel density estimation}
\label{sec:kde}
\begin{figure}[]
\includegraphics[height=7.4cm,width=8.6cm,angle=0]{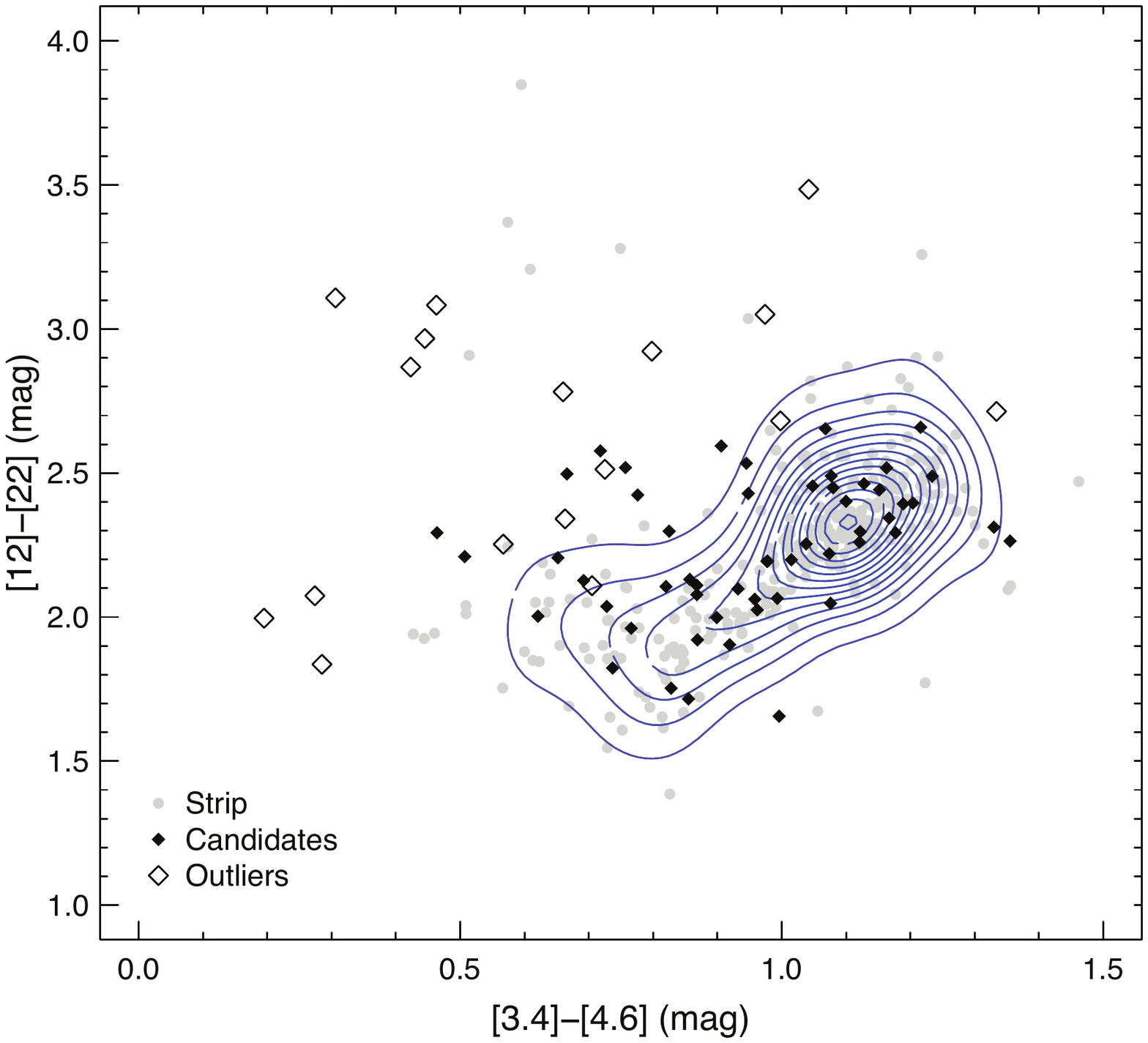}
\caption{The isodensity contours drawn from the KDE technique for the blazars of the \gstrp\ (grey circles) 
are shown for the case of the [3.4]-[4.6]-[12]-[22] $\mu$m 2D projection. 
The AGU identified as blazar candidates (filled black squares) are also shown
in comparison with the outliers (black open square) to show the consistency between the \gstrp\ parametrization  
and the KDE analysis.}
\label{fig:kde_proj3}
\end{figure}

To test our analysis, we also performed a statistical investigation based on an 
independent non-parametric method as the KDE technique as already proposed in Paper I
\citep[see also ][and reference therein]{dabrusco09,laurino11}.
The KDE method provides an effective way of estimating the probability function of a multivariate
distribution and do not require any assumption about the shape of the ``parent" distributions. 
In Figure~\ref{fig:kde_proj3}, 
the isodensity contours drawn from the KDE density probabilities and associated with different levels of density 
are plotted for the blazars of the \gstrp\ in its [3.4]-[4.6]-[12]-[22] $\mu$m 2D projection.

Consequently, for a generic source in the \wse\ archive we can provide an estimate of the probability $\pi_{kde}$that a blazar 
of the \gstrp\ has the same IR colors, this is a surrogate of the probability that a \wse\ source is consistent with the \gstrp.
In Figure~\ref{fig:kde_proj3}, we also show the AGU counterparts with respect to the isodensity contours of the 
\gstrp, to highlight the outliers.
We also report in Table~\ref{tab:outliers} and in Table~\ref{tab:candidates}, the value of $\pi_{kde}$ for each AGU analyzed.

Finally, we note that there are some AGUs for which the KDE analysis suggests that the source is not consistent with the \gstrp, even if the parametric method 
indicates it as a possible candidate. 
The reason for this to happen is that, as previously mentioned, the KDE method does not take 
into account the errors on the IR colors. As a consequence, sources far from the \gstrp\ but 
with large errors could be associated to low density values, as calculated by the KDE method, 
and discarded. However, our parametrization of the \gstrp\ allows us to take 
into account the errors on the \wse\ colors and to consider also this type of sources.
We emphasize that all the sources that the \gstrp\ parametrization indicates as outliers have also $\pi_{kde}$ typically lower than $\sim$ 1\%
of being consistent with the \gstrp.

\section{An analysis of possible selection effects}
\label{sec:effects}
In future, thanks to the developed \gstrp\ parametrization, we will be also able to investigate if there are selection effects that 
could affect our analysis as for example driving the \gstrp\ thickness. At the current stage of our study,
we are able to estimate when a generic \wse\ sources in consistent with the \gstrp\ itself that is a 
necessary tool to compare different samples for future investigations;
Thanks to the parametrization developed we will be also able to verify 
if there are IR blazars that belong to the \gstrp\ but are not detected in $\gamma$-rays, and 
which could be the physical conditions if this occurs.

\begin{figure}[]
\includegraphics[height=6.4cm,width=8.8cm,angle=0]{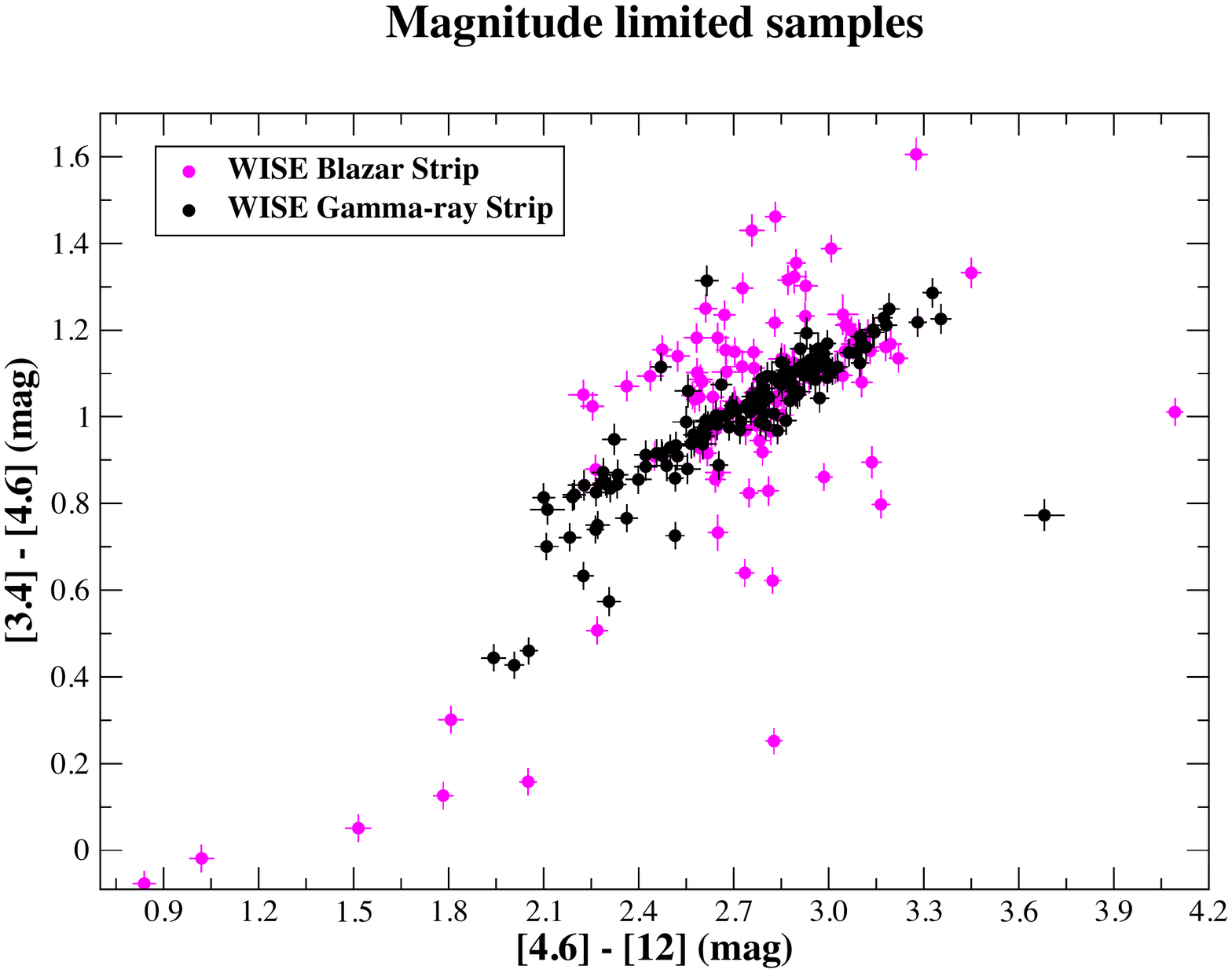}
\caption{The 2D projection of the \wse\ Blazar Strip (magenta)
and the subregion of the \gstrp\ (black) in the IR color diagram [3.4]-[4.6]-[12] $\mu$m
when only bright IR blazars are considered (i.e., those with \wse\ magnitudes in the ranges: $m_1 \leq$13.5, $m_2 \leq$12, $m_3 \leq$11, $m_4 \leq$7.5, respctively).}
\label{fig:limit}
\end{figure}

We remark that the link between the IR and the $\gamma$-ray properties of blazars, is mainly due to the relation between the 
blazar spectral shape in the IR and in the $\gamma$-rays.
To evaluate if selection effects due to flux limits in the selected sample could affect our \gstrp\ description we performed the following tests.
We restricted our analysis to the bright \wse\ blazars with IR magnitudes in the ranges: $m_1 \leq$13.5, $m_2 \leq$12, $m_3 \leq$11, $m_4 \leq$7.5,
that belong to both the \wse\ Blazar strip and to the \gstrp, and we found that a difference in their thickness is still evident (see Figure~\ref{fig:limit}
for the standard 2D projection in [3.4]-[4.6]-[12] $\mu$m color diagram).
This plot suggests that the origin of the \gstrp\ is not due to a selection of bright IR blazars.

\begin{figure}[]
\includegraphics[height=6.4cm,width=8.8cm,angle=0]{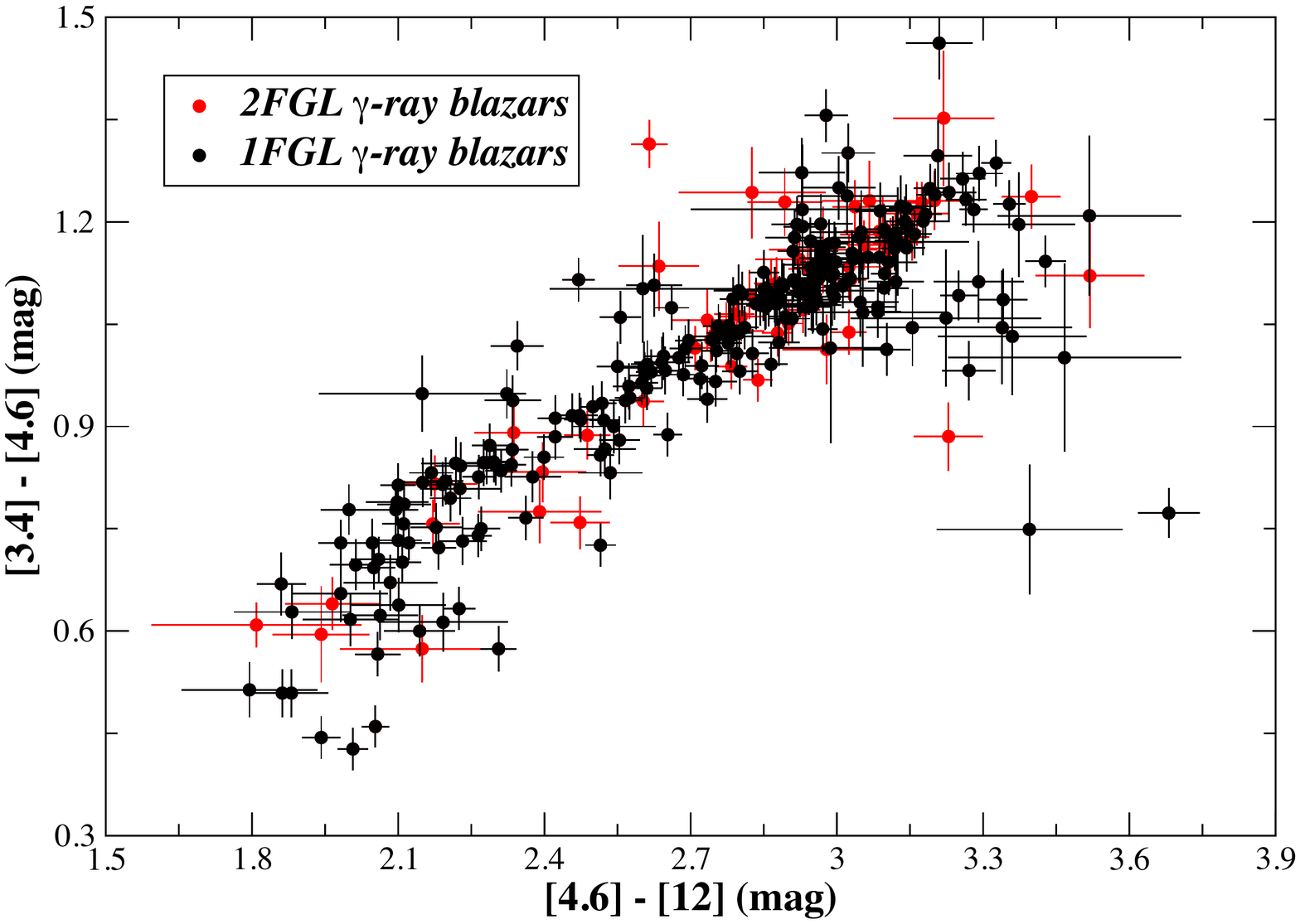}
\caption{The 2D projection of the \gstrp\ in the IR color diagram [3.4]-[4.6]-[12] $\mu$m
when only bright $\gamma$-ray blazars listed in the 1FGL (black) are considered in comparison with those in the 2FGL (red).
There are no clear differences on the thickness of the \gstrp\ in this 2D projection due to the different
samples considered.}
\label{fig:1fgl}
\end{figure}

We also compared the \gstrp\ as formed by the blazars present in the 2FGL and those detected in the first \fer\ LAT catalog (1FGL) \citep{abdo10},
and again we did not find any clear difference between the \gstrp\ drawn with the bright or the faint $\gamma$-ray blazars.
This again suggests that the relation between the \gstrp\ and the $\gamma$-ray detectability is related to the blazar spectral shape
(see Figure~\ref{fig:1fgl} for the standard 2D projection in [3.4]-[4.6]-[12] $\mu$m color diagram).
On the basis of the \gstrp\ parametrization presented in this paper, the above issues will be deeply 
addressed in future works, after the full release of the \wse\ all-sky survey.

\section{Summary and Discussion}
\label{sec:summary}
On the basis of the recent results on the characterization of the IR colors of blazars provided by \wse\ 
(Paper I) and on the comparison with their $\gamma$-ray emission (Paper II),
we developed a method based the \wse\ Blazar Strip 
to identify blazar counterparts of $\gamma$-ray sources.

We developed a method to parametrize the \gstrp\ in the 3D color diagrams based on its 2D projections.
This method is characterized by the use a continuos parameter $s$, in the range 0 -- 1,
that takes into account of the errors of all the IR colors and provides clues on the position of a generic \wse\ source
relative to the \gstrp\ in the 3D color parameter space. High values of the $s$ 
parameters are associated to sources that lie inside the \gstrp\
(as the $\gamma$-ray blazar population of the 2FB sample).
The \gstrp\ has been parametrized in two subregions, the first containing the BZBs while the other with the BZQs 
although in the present work we are only interest in searching for blazar counterparts that lie on the \gstrp.

We applied our parametrization to the sample of the AGUs selected from the 2FGL and the 2LAC.
We found that there are 148 AGUs that can be analyzed
within the footprint of the \wse\ preliminary source catalog.
However, according to our association procedure (see Secton~\ref{sec:associations}) 
only 60 AGUs have a unique \wse\ counterpart without any upper limit on the \wse\ magnitude values.

Then, we calculated the distributions of their $s$ parameter and found that 54 out of 
60 AGUs analyzed are consistent with the \gstrp, corresponding to 
the 90\% of the $\gamma$-ray counterparts analyzed while the remaining 
6 AGU counterparts are outliers of the \gstrp.
In particular, for the 54 AGUs that are consistent with the \gstrp\ we also found that the correlation between the 
$\alpha_{IR}$ and $\alpha_{\gamma}$ is in agreement with that found for the blazars that constitute the \gstrp\ itself
while the same correlation for the 6 outliers is inconsistent with it.

We also applied the KDE non-parametric test to obtain the probability that an AGU 
counterpart belong to the \gstrp\ and we found consistent results with our parametrization (see Section~\ref{sec:kde} for more details).

In addition, an extensive investigation of all the unidentified $\gamma$-ray sources in the 2FGL that fall in the area of the sky where the \wse\ 
preliminary data  have been already released will be provided in a forthcoming paper \citep{massaro12a}. 
Searching for blazar candidates within the unidentified $\gamma$-ray source sample
could potentially leading to the discovery new class of $\gamma$-ray emitting sources.  

Further improvements of the \gstrp\ parametrization will be also possible in the future,
when the whole \wse\ catalog will be available and this parametric method would be calibrated at
different $b$ values not only to look for counterparts of $\gamma$-ray sources but also to search for new blazar candidates
all over the sky \citep{massaro12b}.

\acknowledgements
We are grateful to the anonymous referee for several constructive comments that have been helpful toward improving our presentation.
F. Massaro thank A. Cavaliere, S. Digel, M. Elvis, D. Harris, J. Knodlseder and D. Thompson for their fruitful discussions
and to P. Giommi for his help with the ROMA-BZCAT analysis.
F. Massaro also thanks D. Weedman for his helpful suggestions on the starburst galaxies.
The work at SAO is supported in part by the NASA grant NNX10AD50G and NNX10AD68G.
R. D'Abrusco gratefully acknowledges the financial support of the US Virtual Astronomical Observatory, which is sponsored by the
National Science Foundation and the National Aeronautics and Space Administration.
F. Massaro acknowledges the Fondazione Angelo Della Riccia for the grant awarded him to support 
his research at SAO during 2011 and the Foundation BLANCEFLOR Boncompagni-Ludovisi, n'ee Bildt
for the grant awarded him in 2010 to support his research.
TOPCAT\footnote{\underline{www.star.bris.ac.uk/$\sim$mbt/topcat/}} 
\citep{taylor2005} was used extensively in this work for the preparation and manipulation of the tabular data.
Part of this work is based on archival data, software or on-line services provided by the ASI Science Data Center.
This publication makes use of data products from the Wide-field Infrared Survey Explorer, 
which is a joint project of the University of California, Los Angeles, and the Jet Propulsion Laboratory/California Institute of Technology, 
funded by the National Aeronautics and Space Administration.

{}

\begin{sidewaystable}
\tiny
\caption{The parameters of the \wse\ counterparts for the outliers of the \gstrp.}
\begin{tabular}{|lllcccccccccccc|}
\hline
  \multicolumn{1}{|c}{2FGL name} &
  \multicolumn{1}{c}{\wse\ name} &
  \multicolumn{1}{c}{Name} &
  \multicolumn{1}{c}{Class} &
  \multicolumn{1}{c}{$c_{1}$} &
  \multicolumn{1}{c}{$\sigma_{1}$} &
  \multicolumn{1}{c}{$c_{2}$} &
  \multicolumn{1}{c}{$\sigma_{2}$} &
  \multicolumn{1}{c}{$c_{3}$} &
  \multicolumn{1}{c}{$\sigma_{3}$} &
  \multicolumn{1}{c}{$\alpha_{IR}$} &
  \multicolumn{1}{c}{$\sigma_{\alpha_{IR}}$} &
  \multicolumn{1}{c}{$s_b$} &
  \multicolumn{1}{c}{$s_q$} &
  \multicolumn{1}{c|}{$\pi_{kde}$} \\
\hline
  2FGLJ0532.5-7223 & J053344.71-721623.3 & PMN J0533-7216       & -   & 1.042 & 0.088 & 1.884 & 0.286 & 3.485 & 0.457 & 1.175 & 0.257 & 0.0 & 0.0 & 5.963e-6\\
  2FGLJ0602.3+5315 & J060200.44+531600.2 & GB6 J0601+5315       & HSP & 0.274 & 0.032 & 1.707 & 0.058 & 2.074 & 0.174 & -1.05 & 0.093 & 0.0 & 0.0 & 4.18e-5\\
  2FGLJ0605.0+0001 & J060458.42+000043.2 & GB6 J0604+0000       & -   & 0.285 & 0.034 & 1.218 & 0.089 & 1.836 & 0.431 & -1.02 & 0.099 & 0.0 & 0.0 & 2.09e-9\\
  2FGLJ1304.1-2415 & J130342.56-241442.1 & 1RXS 130343.6-241506 & HSP & 0.195 & 0.041 & 2.506 & 0.073 & 1.996 & 0.269 & -1.28 & 0.118 & 0.0 & 0.0 & 4.49e-9\\
  2FGLJ1753.8-5012 & J175338.55-501513.7 & PMN J1753-5015       & -   & 0.663 & 0.050 & 2.647 & 0.057 & 2.341 & 0.122 & 0.071 & 0.147 & 0.0 & 0.0 & 0.003\\
  2FGLJ1936.9+8402 & J193739.76+835628.9 & 6C B194425+834912    & -   & 1.334 & 0.147 & 2.938 & 0.232 & 2.714 & 0.581 & 2.030 & 0.431 & 0.0 & 0.06 & 0.01\\
\hline
\end{tabular}
\label{tab:outliers}
~\\
Col. (1) Source name reported in the 2FGL \citep{abdo11}.\\
Col. (2) Source name reported in the \wse\ preliminary data release.\\
Col. (3) Source name as reported in the 2LAC \citep{ackermann11}.\\
Col. (4) Source class as reported in the 2LAC: high-synchrotron-peaked blazar (HSP), 
intermediated-synchrotron-peaked blazar (ISP), intermediated-synchrotron-peaked blazar (LSP).\\
Col. (5) IR color $c_1$ = [3.4]-[4.6] $\mu$m.\\
Col. (6) Error $\sigma_1$ on $c_1$.\\
Col. (7) IR color $c_2$ = [4.6]-[12] $\mu$m.\\
Col. (8) Error $\sigma_2$ on $c_1$. \\
Col. (9) IR color $c_3$ = [12]-[22] $\mu$m.\\
Col. (10) Error $\sigma_3$ on $c_3$. \\
Col. (11) IR spectral index $\alpha_{IR}$.\\
Col. (12) Error on $\alpha_{IR}$. \\
Col. (13,14) $s_b$ and $s_q$ values.\\
Col. (15) Probability derived from the KDE analysis $\pi_{kde}$.\\
\end{sidewaystable}

\begin{sidewaystable}
\tiny
\caption{The parameters of the \wse\ counterparts for the blazar candidates.}
\begin{tabular}{|lllcccccccccccc|}
\hline
  \multicolumn{1}{|c}{2FGL name} &
  \multicolumn{1}{c}{\wse\ name} &
  \multicolumn{1}{c}{Name} &
  \multicolumn{1}{c}{Class} &
  \multicolumn{1}{c}{$c_{1}$} &
  \multicolumn{1}{c}{$\sigma_{1}$} &
  \multicolumn{1}{c}{$c_{2}$} &
  \multicolumn{1}{c}{$\sigma_{2}$} &
  \multicolumn{1}{c}{$c_{3}$} &
  \multicolumn{1}{c}{$\sigma_{3}$} &
  \multicolumn{1}{c}{$\alpha_{IR}$} &
  \multicolumn{1}{c}{$\sigma_{\alpha_{IR}}$} &
  \multicolumn{1}{c}{$s_b$} &
  \multicolumn{1}{c}{$s_q$} &
  \multicolumn{1}{c|}{$\pi_{kde}$} \\
\hline
  2FGLJ0156.4+3909 & J015631.40+391430.5 & MG4 J015630+3913     & -   & 1.167 & 0.042 & 2.731 & 0.063 & 2.344 & 0.196 & 1.541 & 0.124 & 0.273 & 0.355 & 0.28  \\
  2FGLJ0248.6+8440 & J024948.29+843556.9 & NVSS J024948+843556  & -   & 0.958 & 0.036 & 2.654 & 0.039 & 2.062 & 0.098 & 0.927 & 0.105 & 0.548 & 0.475 & 0.41  \\
  2FGLJ0253.4+3218 & J025333.64+321720.4 & MG3 J025334+3217     & -   & 1.163 & 0.043 & 2.954 & 0.060 & 2.518 & 0.117 & 1.529 & 0.126 & 0.317 & 0.428 & 0.60  \\
  2FGLJ0309.3-0743 & J030943.22-074427.5 & NVSS J030943-074427  & HSP & 0.728 & 0.036 & 2.128 & 0.071 & 2.037 & 0.248 & 0.260 & 0.106 & 0.327 & 0.109 & 0.18  \\
  2FGLJ0332.5-1118 & J033223.25-111950.6 & NVSS J033223-111951  & HSP & 0.977 & 0.033 & 2.652 & 0.033 & 2.194 & 0.050 & 0.986 & 0.099 & 0.739 & 0.748 & 0.43  \\
  2FGLJ0333.7+2918 & J033349.00+291631.6 & TXS 0330+291         & ISP & 0.820 & 0.036 & 2.096 & 0.048 & 2.106 & 0.134 & 0.525 & 0.105 & 0.460 & 0.0   & 0.09  \\
  2FGLJ0424.3-5332 & J042504.26-533158.3 & PMN J0425-5331       & ISP & 0.993 & 0.032 & 2.614 & 0.031 & 2.065 & 0.050 & 1.032 & 0.095 & 0.762 & 0.733 & 0.41  \\
  2FGLJ0438.8-4521 & J043900.84-452222.2 & PKS 0437-454         & LSP & 1.189 & 0.037 & 3.111 & 0.041 & 2.393 & 0.087 & 1.605 & 0.109 & 0.287 & 0.559 & 0.74  \\
  2FGLJ0440.1-3211 & J043933.88-321009.8 & PKS 0437-322         & LSP & 0.757 & 0.069 & 2.701 & 0.148 & 2.519 & 0.422 & 0.344 & 0.201 & 0.114 & 0.078 & 4.12e-4\\
  2FGLJ0440.4+1433 & J044021.13+143756.5 & TXS 0437+145         & -   & 1.234 & 0.048 & 3.020 & 0.068 & 2.489 & 0.163 & 1.737 & 0.140 & 0.0   & 0.354 & 0.43  \\
  2FGLJ0456.5+2658 & J045617.36+270220.7 & MG2 J045613+2702     & -   & 1.122 & 0.043 & 2.903 & 0.061 & 2.297 & 0.159 & 1.409 & 0.128 & 0.377 & 0.381 & 1.00  \\
  2FGLJ0505.9+6116 & J050558.78+611335.6 & NVSS J050558+611336  & HSP & 0.766 & 0.041 & 1.948 & 0.085 & 1.961 & 0.347 & 0.371 & 0.119 & 0.211 & 0.0   & 0.11  \\
  2FGLJ0508.1-1936 & J050818.99-193556.0 & PMN J0508-1936       & -   & 1.048 & 0.070 & 2.901 & 0.151 & 2.455 & 0.474 & 1.190 & 0.205 & 0.128 & 0.150 & 0.59  \\
  2FGLJ0525.5-6011 & J052542.42-601340.8 & SUMSS J052542-601341 & HSP & 0.718 & 0.044 & 2.268 & 0.125 & 2.577 & 0.369 & 0.231 & 0.129 & 0.163 & 0.093 & 0.01  \\
  2FGLJ0526.8+6326 & J052606.71+631729.0 & GB6 J0526+6317       & -   & 0.996 & 0.045 & 2.709 & 0.077 & 1.656 & 0.349 & 1.041 & 0.134 & 0.193 & 0.140 & 0.02  \\
  2FGLJ0532.0-4826 & J053158.61-482736.1 & PMN J0531-4827       & LSP & 0.899 & 0.031 & 2.492 & 0.029 & 1.998 & 0.039 & 0.755 & 0.090 & 0.862 & 0.457 & 0.37  \\
  2FGLJ0537.7-5716 & J053748.95-571830.0 & SUMSS J053748-571828 & ISP & 0.857 & 0.033 & 2.348 & 0.041 & 2.131 & 0.112 & 0.636 & 0.096 & 0.530 & 0.281 & 0.17  \\
  2FGLJ0538.5-3909 & J053810.35-390842.5 & 1RXS 053810.0-390839 & HSP & 0.621 & 0.041 & 2.032 & 0.118 & 2.003 & 0.492 &-0.050 & 0.119 & 0.194 & 0.061 & 0.13  \\
  2FGLJ0604.2-4817 & J060408.61-481724.9 & 1ES 0602-482         & HSP & 0.666 & 0.036 & 1.943 & 0.085 & 2.497 & 0.224 & 0.080 & 0.106 & 0.233 & 0.0   & 0.006 \\
  2FGLJ0609.4-0248 & J060915.06-024754.6 & NVSS J060915-024754  & HSP & 0.855 & 0.036 & 2.218 & 0.064 & 1.716 & 0.321 & 0.627 & 0.107 & 0.284 & 0.130 & 0.21  \\
  2FGLJ0621.9+3750 & J062157.63+375057.4 & GB6 J0621+3750       & -   & 1.204 & 0.053 & 3.096 & 0.070 & 2.396 & 0.161 & 1.646 & 0.155 & 0.164 & 0.339 & 0.68  \\
  2FGLJ0703.1-3912 & J070312.64-391418.7 & NVSS J070312-391418  & ISP & 0.962 & 0.036 & 2.602 & 0.04  & 2.025 & 0.105 & 0.942 & 0.105 & 0.531 & 0.436 & 0.40  \\
  2FGLJ0706.5+7741 & J070651.32+774137.1 & NVSS J070651+774137  & ISP & 0.919 & 0.033 & 2.489 & 0.034 & 1.904 & 0.079 & 0.816 & 0.097 & 0.628 & 0.264 & 0.31  \\
  2FGLJ0734.2-7706 & J073443.42-771114.1 & PKS 0736-770         & LSP & 1.128 & 0.041 & 3.165 & 0.052 & 2.463 & 0.134 & 1.427 & 0.120 & 0.334 & 0.435 & 0.59  \\
  2FGLJ0823.0+4041 & J082257.55+404149.8 & B3 0819+408          & LSP & 1.177 & 0.037 & 2.963 & 0.041 & 2.293 & 0.072 & 1.570 & 0.109 & 0.512 & 0.596 & 0.70  \\
  2FGLJ0856.0+7136 & J085654.85+714623.4 & GB6 J0856+7146       & LSP & 1.152 & 0.036 & 3.007 & 0.04  & 2.442 & 0.071 & 1.494 & 0.107 & 0.450 & 0.608 & 0.88  \\
  2FGLJ1021.6+8021 & J102202.09+802349.9 & WN B1016.6+8038      & -   & 0.945 & 0.046 & 2.741 & 0.083 & 2.534 & 0.212 & 0.889 & 0.134 & 0.278 & 0.263 & 0.06  \\
  2FGLJ1029.9+7437 & J103122.15+744157.0 & S5 1027+74           & ISP & 0.507 & 0.032 & 2.269 & 0.035 & 2.210 & 0.063 &-0.383 & 0.094 & 0.387 & 0.0   & 0.01  \\
  2FGLJ1304.3-4353 & J130421.02-435310.1 & 1RXS 130421.2-435308 & HSP & 0.869 & 0.034 & 2.268 & 0.032 & 1.921 & 0.053 & 0.671 & 0.099 & 0.727 & 0.305 & 0.30  \\
  2FGLJ1351.3-2909 & J135146.86-291217.5 & PKS 1348-289         & -   & 1.077 & 0.040 & 2.885 & 0.059 & 2.49  & 0.137 & 1.275 & 0.117 & 0.392 & 0.415 & 0.62  \\
  2FGLJ1406.2-2510 & J140609.61-250808.6 & NVSS J140609-250808  & -   & 0.868 & 0.040 & 2.461 & 0.067 & 2.110 & 0.228 & 0.668 & 0.117 & 0.333 & 0.209 & 0.24  \\
  2FGLJ1407.5-4257 & J140739.73-430231.9 & CGRaBS J1407-4302    & LSP & 0.948 & 0.043 & 2.724 & 0.066 & 2.429 & 0.149 & 0.901 & 0.125 & 0.377 & 0.365 & 0.12  \\
  2FGLJ1416.0+1323 & J141558.82+132023.8 & PKS B1413+135        & LSP & 1.355 & 0.031 & 2.897 & 0.030 & 2.264 & 0.038 & 2.089 & 0.093 & 0.0   & 0.766 & 0.03  \\
  2FGLJ1416.3-2415 & J141612.18-241813.4 & NVSS J141612-241812  & HSP & 0.464 & 0.036 & 1.702 & 0.093 & 2.293 & 0.277 &-0.508 & 0.106 & 0.141 & 0.0   & 0.003 \\
  2FGLJ1419.4+7730 & J141900.31+773228.8 & 1RXS 141901.8+773229 & HSP & 0.652 & 0.041 & 2.013 & 0.116 & 2.206 & 0.493 & 0.039 & 0.119 & 0.195 & 0.0   & 0.09  \\
  2FGLJ1419.4-0835 & J141922.56-083832.1 & NVSS J141922-083830  & LSP & 1.121 & 0.037 & 2.885 & 0.045 & 2.26  & 0.091 & 1.406 & 0.109 & 0.528 & 0.534 & 0.92  \\
  2FGLJ1421.1-1117 & J142100.15-111820.1 & PMN J1420-1118       & -   & 1.080 & 0.056 & 2.714 & 0.122 & 2.449 & 0.318 & 1.286 & 0.165 & 0.208 & 0.207 & 0.39  \\
  2FGLJ1518.2-2733 & J151803.60-273131.0 & TXS 1515-273         & -   & 0.692 & 0.033 & 2.190 & 0.044 & 2.127 & 0.097 & 0.153 & 0.098 & 0.539 & 0.180 & 0.11  \\
  2FGLJ1553.2-2424 & J155331.63-242205.7 & PKS 1550-242         & -   & 1.100 & 0.037 & 3.058 & 0.045 & 2.402 & 0.090 & 1.342 & 0.109 & 0.506 & 0.536 & 0.85  \\
  2FGLJ1558.3+8513 & J160031.78+850949.2 & WN B1609.6+8517      & LSP & 0.868 & 0.034 & 2.301 & 0.043 & 2.078 & 0.132 & 0.668 & 0.100 & 0.487 & 0.234 & 0.21  \\
  2FGLJ1558.6-7039 & J155736.14-704027.1 & PKS 1552-705         & -   & 0.906 & 0.082 & 3.042 & 0.151 & 2.594 & 0.354 & 0.776 & 0.241 & 0.0   & 0.142 & 0.01  \\
  2FGLJ1725.1-7714 & J172350.85-771349.9 & PKS 1716-771         & -   & 1.038 & 0.033 & 2.769 & 0.040 & 2.254 & 0.092 & 1.164 & 0.097 & 0.568 & 0.575 & 0.72  \\
  2FGLJ1759.2-4819 & J175858.45-482112.6 & PMN J1758-4820       & -   & 0.978 & 0.038 & 2.609 & 0.036 & 2.194 & 0.057 & 0.986 & 0.112 & 0.655 & 0.635 & 0.39  \\
  2FGLJ1816.7-4942 & J181656.00-494344.0 & PMN J1816-4943       & -   & 1.076 & 0.051 & 2.868 & 0.066 & 2.048 & 0.213 & 1.272 & 0.149 & 0.274 & 0.320 & 0.32  \\
  2FGLJ1818.7+2138 & J181905.22+213234.0 & MG2 J181902+2132     & -   & 0.932 & 0.038 & 2.464 & 0.055 & 2.098 & 0.187 & 0.851 & 0.111 & 0.386 & 0.261 & 0.31  \\
  2FGLJ1825.1-5231 & J182513.82-523057.7 & PKS 1821-525         & -   & 1.074 & 0.034 & 2.864 & 0.034 & 2.220 & 0.067 & 1.269 & 0.099 & 0.658 & 0.666 & 0.87  \\
  2FGLJ1829.1+2725 & J182913.97+272902.9 & 87GB 182712.0+272717 & -   & 0.776 & 0.042 & 2.796 & 0.059 & 2.424 & 0.157 & 0.400 & 0.123 & 0.0   & 0.147 & 8.65e-4\\ 
  2FGLJ1830.0+1325 & J183000.76+132414.4 & MG1 J183001+1323     & -   & 1.015 & 0.043 & 2.631 & 0.047 & 2.199 & 0.122 & 1.094 & 0.126 & 0.451 & 0.457 & 0.46  \\
  2FGLJ1858.1-2510 & J185819.07-251050.5 & PMN J1858-2511       & -   & 1.068 & 0.106 & 3.183 & 0.100 & 2.654 & 0.179 & 1.248 & 0.310 & 0.123 & 0.231 & 0.14  \\
  2FGLJ1913.4+4440 & J191401.88+443832.5 & 1RXS 191401.9+443849 & HSP & 0.737 & 0.039 & 2.256 & 0.081 & 1.823 & 0.400 & 0.283 & 0.115 & 0.249 & 0.125 & 0.20  \\
  2FGLJ1940.8-6213 & J194121.77-621120.8 & PKS 1936-623         & -   & 1.330 & 0.036 & 2.860 & 0.043 & 2.312 & 0.092 & 2.018 & 0.107 & 0.0   & 0.542 & 0.04  \\
  2FGLJ1941.6+7218 & J194127.01+722142.2 & 87GB 194202.1+721428 & -   & 1.216 & 0.080 & 3.383 & 0.124 & 2.659 & 0.271 & 1.684 & 0.236 & 0.0   & 0.194 & 0.13  \\
  2FGLJ1959.6-2931 & J200016.96-293026.2 & PMN J2000-2931       & -   & 0.825 & 0.049 & 2.398 & 0.094 & 2.298 & 0.304 & 0.539 & 0.144 & 0.241 & 0.147 & 0.04  \\
  2FGLJ2103.6-6236 & J210338.37-623225.5 & PMN J2103-6232       & HSP & 0.828 & 0.033 & 2.312 & 0.037 & 1.753 & 0.132 & 0.548 & 0.096 & 0.516 & 0.217 & 0.26  \\
\hline
\end{tabular}
\label{tab:candidates}
~\\
Col. (1) Source name reported in the 2FGL \citep{abdo11}.\\
Col. (2) Source name reported in the \wse\ preliminary data release.\\
Col. (3) Source name as reported in the 2LAC \citep{ackermann11}.\\
Col. (4) Source class as reported in the 2LAC: high-synchrotron-peaked blazar (HSP), 
intermediated-synchrotron-peaked blazar (ISP), intermediated-synchrotron-peaked blazar (LSP).\\
Col. (5) IR color $c_1$ = [3.4]-[4.6] $\mu$m.\\
Col. (6) Error $\sigma_1$ on $c_1$.\\
Col. (7) IR color $c_2$ = [4.6]-[12] $\mu$m.\\
Col. (8) Error $\sigma_2$ on $c_1$. \\
Col. (9) IR color $c_3$ = [12]-[22] $\mu$m.\\
Col. (10) Error $\sigma_3$ on $c_3$. \\
Col. (11) IR spectral index $\alpha_{IR}$.\\
Col. (12) Error on $\alpha_{IR}$. \\
Col. (13,14) $s_b$ and $s_q$ values.\\
Col. (15) Probability derived from the KDE analysis $\pi_{kde}$.\\
\end{sidewaystable}

\end{document}